\documentstyle[12pt]{article}
\begin{document}
\newcommand{\ket}[1] {\mbox{$ \vert #1 \rangle $}}
\newcommand{\bra}[1] {\mbox{$ \langle #1 \vert $}}
\newcommand{\bkn}[1] {\mbox{$ < #1 > $}}
\newcommand{\bk}[1] {\mbox{$ \langle #1 \rangle $}}
\newcommand{\scal}[2]{\mbox{$ \langle #1 \vert #2  \rangle $}}
\newcommand{\expect}[3] {\mbox{$ \bra{#1} #2 \ket{#3} $}}
\newcommand{\ki}{\mbox{$ \ket{\psi_i} $}}
\newcommand{\bi}{\mbox{$ \bra{\psi_i} $}}
\newcommand{\p} \prime
\newcommand{\e} \epsilon
\newcommand{\la} \lambda
\newcommand{\om} \omega   \newcommand{\Om} \Omega
\newcommand{\cc}{\mbox{$\cal C $}}
\newcommand{\w} {\hbox{ weak }}
\newcommand{\al} \alpha
\newcommand{\bt} \beta

\newcommand{\be} {\begin{equation}}
\newcommand{\ee} {\end{equation}}
\newcommand{\ba} {\begin{eqnarray}}
\newcommand{\ea} {\end{eqnarray}}

\def\lrD{\mathrel{{\cal D}\kern-1.em\raise1.75ex\hbox{$\leftrightarrow$}}}
\def\lr #1{\mathrel{#1\kern-1.25em\raise1.75ex\hbox{$\leftrightarrow$}}}
%i\! \lr{\partial_V}

\overfullrule=0pt \def\sqr#1#2{{\vcenter{\vbox{\hrule height.#2pt
          \hbox{\vrule width.#2pt height#1pt \kern#1pt
           \vrule width.#2pt}
           \hrule height.#2pt}}}}
\def\square{\mathchoice\sqr68\sqr68\sqr{4.2}6\sqr{3}6} \def\lrpartial{\mathrel
{\partial\kern-.75em\raise1.75ex\hbox{$\leftrightarrow$}}}

\begin{flushright}
LPTENS 96/46\\
october 1996
\end{flushright}
\vskip 1. truecm
\vskip 1. truecm
\centerline{\LARGE\bf{Time dependent perturbation theory}}
\vskip 2 truemm
%\centerline{\LARGE\bf{from Wheeler De Witt solutions}}
\centerline{\LARGE\bf{in Quantum Cosmology}}
\vskip 1. truecm
\vskip 1. truecm

\centerline{{\bf R. Parentani}\footnote{present address: 
Laboratoire de Math\'ematique et Physique Th\'eorique,
Facult\' e des Sciences,
\newline
${ \quad\quad\quad
\quad\quad\quad
\quad\quad\quad } $  Universit\'e de Tours, 37200 Tours, France. ${ \ }$
%\newline
%e-mail: parenta@celfi.phys.univ-tours.fr
}}

\vskip 5 truemm
\centerline{Laboratoire de Physique Th\'eorique de l'\' Ecole
Normale Sup\'erieure\footnote{Unit\'e propre de recherche du C.N.R.S.
associ\'ee \`a l'ENS et \`a l'Universit\'e de Paris Sud.}}
%Ecole
%Normale Sup\'erieure
\centerline{24 rue Lhomond,
75.231 Paris CEDEX 05, France.}
\centerline{e-mail: parenta@celfi.phys.univ-tours.fr}
\vskip 5 truemm
\vskip 1.5 truecm

\vskip 1.5 truecm
\vskip 1. truecm
%\vskip 1.5 truecm
{\bf Abstract }
We describe radiative 
processes in Quantum Cosmology,
%. To this end, we construct 
%time-dependent perturbation theory 
from the solutions of the Wheeler De Witt equation.
By virtue of this constraint equation,
%the momentum operator of the cosmological expansion
the quantum propagation of gravity is modified by the 
% gravity automatically incorporates the 
matter interaction
hamiltonian at the level of amplitudes.
In this we generalize previous works where gravity was 
coupled only to expectation values
of matter operators.
% were coupled to
%used to modify the propagation of gravity. 
%As a byproduct of From this extension,
% By examining those amplitudes
%we show how 
%the concept of conditional 
By a ``reduction formula'' we show how to obtain  
 transition amplitudes from the entangled gravity+matter system.
%Furthermore, we determine under which conditions
% the usual time-dependent 
%perturbation theory is recovered.  
%not only for expectation values thereby generalizing the 
%notion of backreaction at the quantum level.
Then we show how {\it each} transition among
matter constituents of the universe determines dynamically
 {\it its} background from which a time 
parameter is defined.
%The validity of the use of this inner time parametrization 
%is discussed in physical terms.
Finally, we leave the mini-superspace context
by introducing an extended formalism
in which the momenta of the exchanged quanta
no longer vanish. Then, the concept of
spatial displacement 
%associated with a time lapse 
emerges from radiative processes
like the time parametrization did, thereby
unifying the way by which  space and time intervals
 are recovered in quantum cosmology.

\vfill \newpage

\section{Introduction}

At present quantum gravity does not exist.
However, we know that gravity induces quantum effects.
Particularly impressive examples are black hole evaporation\cite{hawk} 
and pair creation of massive particles in a cosmological situation\cite{parker}.
Both results are derived by treating gravity at a classical
level. Only the matter fields propagating in the background
geometry fluctuate quantum mechanically\cite{BD}.   This asymmetric treatment
raises immediately  the problem of how to take into account the 
backreaction of these quanta. Indeed, by maintaining gravity classical,
the backreaction can only be approximated by the semi-classical 
treatment\cite{wald}
in which the expectation value of the energy momentum tensor acts a
source of the Einstein equations, all fluctuations being 
ignored by gravity.

Besides the neglection of the corrections induced by gravitational loops, 
this asymmetric treatment
%finds probably its justification in the fact 
%that the gravitational quantum corrections are controlled by the Planck mass,
% concerning the ``transPlanckian'' frequencies
%which occur in those spontaneous pair creation processes).
%Besides these quantum corrections, the asymmetric treatment 
also relies on the hypothesis that gravity might be treated
in the background field approximation (BFA), i.e. gravity is described
by a {\it single} classical geometry which is {\it insensitive} to
the individual quantum matter events it engenders.
%The question of the validity of this approximation is the 
%main concern of this paper. 
In this respect, we recall the
%at there is a
controversial\cite{wald}\cite{thooft}\cite{jacobs}
%\cite{THooft2}\cite{emp}\cite{MaPa}
question concerning the role of the ``transPlanckian'' frequencies occurring
 in black hole evaporation:
% The controversial problem is the following:
In the presence of these high frequencies, is it 
still meaningful to treat gravity at the BFA as supposed in \cite{hawk}
or is the geometrical background also
determined\cite{THooft2}--\cite{itz}
%\cite{emp}\cite{MaPa}\cite{fe}\cite{itz}
by the final matter state ?
%Can the The question  in those spontaneous pair creation processes)

The main concern of this paper is to address the question of 
the validity of the
BFA and to determine the conceptual and quantitative
changes when this approximation is no longer valid. 
As a corollary, we shall
show that the question of this validity is directly related to 
the so-called ``problem of time''
in quantum cosmology, see for reviews 
 \cite{isham}\cite{hartle}\cite{ontime}\cite{kiefer2}.
%In particular,
%% we shall show that 
%the analysis
%of Green functions performed in \cite{wdwgf} directly 
%applies to the present dynamical treatment.

 The determination of the validity of the background field approximation
%necessitates 
requires the construction of a more quantum mechanical framework.
Then, one should choose some physical processes
and describe them
% described by transition amplitudes
%must be evaluated 
in both frameworks, i.e. at the BFA and in the new framework. 
From  both expressions
of the transition amplitude of the same process,
one can determine under which circumstances
%, whether or not
%the two versions of the transition amplitude of the same process 
the two versions coincide. In a first attempt to question
the validity of the BFA, this approach 
has been applied in \cite{rec}\cite{recmir}\cite{recR}\cite{recV}
 to accelerated systems. 
In this case, what plays the role of the background is the classical 
accelerated trajectory which is replaced by a dynamically determined
wave function in the more quantum mechanical framework. 

In this paper, we apply the same techniques to quantum cosmology.
In quantum cosmology, 
%this approximation 
%is no longer used since 
the gravitational degrees of freedom
are quantized, at least in some restricted domain\footnote{
One may put into question the physical validity of such a restriction.
As a first answer, we recall that there is a large variety of
physical situations in which it is relevant to quantize the fluctuations
of a very restricted number of variables, c.f. the notion 
of quasi-particle. Furthermore, quantum mechanics was not
introduced to explain the Lamb shift effect but rather to deal with the
Compton effect, that is a tree level effect.
It is therefore not too unreasonable to hope that there is 
some physical relevance in dealing with mini-superspace.} 
such as mini-superspace\cite{HH} or to a
 quadratic approximation as in \cite{HalH}.
We use the mini-superspace framework to question the validity
of the BFA by considering processes
%problem of how to perturbatively include 
%interactions 
among  matter constituents of the
universe. By obtaining explicit expressions for the transition amplitudes
describing these processes
and comparing them with the corresponding amplitudes evaluated in the
BFA,
we shall be in position to display, both conceptually and quantitatively, 
the modifications induced by the quantum dynamical
character of gravity.  

The first task is to define 
the amplitudes of transition in quantum cosmology. 
Indeed, the presence of the constraint, which guarantees that the
theory is invariant under reparametrization of time, implies that 
the interaction hamiltonian of the transition under investigation
must be treated from the ``inside''. Therefore it should be
included in the matter hamiltonian {\it before}
one imposes the constraint, i.e. the Wheeler De Witt (WDW) equation.
%to be satisfied. 
Indeed had one work from the ``outside'', one would have left the sheet of 
zero energy solutions and therefore broken the reparametrization
invariance, see the restrictions mentioned in ref. \cite{gang}.
% when one acts from the outside on a constrained system.

By first including the interaction hamiltonian in the WDW equation and 
then developing the evolution operator of matter+gravity configurations
in power of this hamiltonian, we show how to obtain
transition amplitudes from the correlated quantum propagation
of gravity and matter.
%in quantum cosmology.
Because of the dynamical character of gravity,
%we shall see that these amplitudes cannot be interpreted as amplitudes 
%of probability. We shall therefore introduce conditional
%amplitudes\cite{ah}\cite{GO}
%by removing the contribution of the free propagation of the 
%initial state, where free is defined with respect to the matter 
%interaction hamiltonian.
%some care is required to properly identify
%These conditional amplitudes are 
a ``reduction formula''
sould be applied in order to 
obtain amplitudes of {\it probability}. 
%from which transition rates can be deduced.
%Once this is done,
To first order in the interaction hamiltonian,
 these amputated amplitudes
are equal, in phase and in amplitude, 
to the corresponding amplitudes evaluated at the BFA when one performs
the double development presented in \cite{wdwgf} which consists
in (i) using a WKB expression for the ``free'' propagation 
of matter and gravity
% kernels $K$
%, see eq. (\ref{M16}),
and
(ii) making a
first order development in the matter energy changes induced by the
transition itself.
As an important byproduct of this,
% double development, 
we shall 
see that a time parametrization is 
%automatically 
delivered by
the very fact that we are considering a {\it change} in the matter state.
In this we differ from previous work. 
%The time parametrization appears
%in our transition amplitude from the evaluation of the amplitude itself.
Neither a factorization\cite{HalH}\cite{banks}\cite{BV} of the wave 
function of the universe
 nor a development in $1/G$\cite{kiefer2}\cite{ortiz} must be 
imposed from the outset ($G$ is Newton's constant).

We shall then evaluate the corrections to these approximations.
%These corrections determine under which conditions can one
%use the amplitudes evaluated at the BFA.
%background field approximation.
The origin of these corrections terms lies in the non-linearity of the
propagation of the combined matter+gravity system. 
As advocated in \cite{THooft2}\cite{emp}\cite{wdwgf}, 
we shall see that the background is indeed
determined symmetrically by both the initial and the final matter states.

%In this 
%we differ from previous work based on a
%Born -Oppenheimer development since only (real) expectation 
%values of the matter hamiltonian 
%entered into the gravitational part of the wave function
%of the universe, see {\it e.g.} \cite{HalH}\cite{Hal}\cite{BV}\cite{Panm}.

We have organized the paper as follows.
We first compute the amplitudes of transition at the
%background field approximation (BFA) 
BFA. In this framework, the universe's evolution
is insensitive to dynamical processes and once for all
specified by a classical law $a= a(t)$.
In Section 3, we define and compute the amplitudes 
of the same transitions in quantum cosmology 
from the correlated propagation of gravity and matter.
%in the WDW framework. 
By proceeding to the double development mentioned above,
%as we did in section 2.3., 
we show how to recover the amplitudes evaluated at the BFA.
%in the {\it mean }
%universe in which the interaction occurs.
%The corrections to this double development are considered in Section 4.
In Section 4, an extension in which the exchanged momenta 
no longer vanish is presented. It is shown how the notion of spatial displacement 
emerges from those transitions.
The corrections to the double development are considered in Section 5
together with some comments on quantum gravity.
%In the last Section, we make a resume and make some conjectures

\section{The amplitudes evaluated 
using the Background Field Approximation}

In this section we evaluate 
%y amplitudes evaluated using the
%Background Field Approximation (BFA),
%we mean that the 
matter transition amplitudes in a cosmological
background whose
%do not affect the 
scale factor follows
a given trajectory $a = a(t)$.
By ``Background Field Approximation'' (BFA)
we mean that the classical trajectory $a = a(t)$
is not affected by the energy of the quanta under
examination, hence it is external to the dynamics.
The purpose of the next Section is to treat 
matter interactions and 
gravitational propagation on a same
quantum footing. This will be done by
studying matter interactions from the 
solutions of the Wheeler De Witt equation.
%The validity of this approximation
% which is not modified by the quantum transitions one studies.

The particular transitions we shall study  
are the ones of a massive two field detector coupled 
to a massless field $\phi$. This model was first 
used by Unruh\cite{unruh} to establish the thermalization 
of accelerated systems in vacuum.
Moreover, it has been found useful to investigate the
recoils effects of these systems when one no longer
treats their trajectory as given and classical, see 
\cite{rec}\cite{recR}.
%\cite{recmir}\cite{recV}.
This enlargement of the quantum dynamics bears many
similarities with 
%what shall be done  upon considering
the corresponding one in cosmology,
%solutions of the WDW equation.
see Sections 3 and 4.

In this Section and in Section 3, the simplified detector model
 consists of two homogeneous 
massive fields $\psi_M$ and $\psi_m$
of respective masses $M$ and $m$ which satisfy 
$\Delta m = M - m <\!\!< M$.
%As in \cite{wdwgf}, eq. (\ref{M1}), 
Furthermore, we suppose that $\dot a / a <\!\!< M$ so as
to guarantee that the number of massive fields $n_M + n_m $
%= N$
is conserved. However we take $\dot a / a \simeq \Delta m$
in order to have non vanishing amplitudes for
the spontaneous excitations engendered by
the cosmological expansion. 
%In \cite{wdwpc}, we shall relax the 
%condition  $\dot a / a << M$ and compute the amplitudes of 
%pair creation of massive quanta in quantum cosmology. 
%
As a specific example, we shall compute the amplitudes in de Sitter space
 (the calculation is presented in Appendix A). We have chosen
de Sitter space to obtain the notion of a constant rate 
of transition in a 
%cosmological 
time dependent background and to make
the analogy between accelerated detectors
and inertial detectors in an evolving universe even closer.
%we shall use again 
%exactly this two-field detector model 
%of this last reference in a (almost) de Sitter universe. 
Indeed, in de Sitter space, there are also thermal effects
associated with a static event horizon\cite{HH2}. 
In Section 4, the comparison will be further extended
by considering fields with non-vanishing momenta 
and by taking into account 
 the recoils of the massive detector.
%, recoil
%effects in a cosmological 
%situation will compared with the corresponding
%ones in a external electric field\cite{rec}.
% and in a cosmological 
%situation will be made explicit in Section 4 upon taking into account 
% the recoils of the massive detector.

Throughout the text, we shall use the 
notations and the results of ref. \cite{wdwgf}. 
The reader is invited to consult that paper.

In the proper time gauge, wherein the lapse $N$ is equal to 
one, the free
hamiltonian of the three (rescaled, see \cite{wdwgf}) fields is
 the sum of three harmonic oscillators
\be
 H_{0}(t) =
M d_M^\dagger d_M + m d_m^\dagger d_m + { k \over a(t) } 
 d_\gamma^\dagger d_\gamma 
%g \left[ \psi_M^\dagger \psi_m+
%\psi_M \psi_m^\dagger 
%\right] \phi
%g { (d_M^\dagger d_m + d_m^\dagger d_M )
%(d_\gamma^\dagger +  d_\gamma ) \over {\sqrt{ 2M 2 m 2 k}}}
\label{blu}
\ee
We have introduced the destruction and creator operators of
the three fields and remove the zero point energy. Notice 
that the frequency of the conformal massless $\phi$ field
depends of $a(t)$. In fact this field has a fixed conformal frequency $k$,
where the conformal time is defined by $\eta = \int^t dt/a(t)$.

The hamiltonian interaction which
relates the three (rescaled) fields is 
\be
H_{int}(t) = g \left[ \psi_M^\dagger \psi_m+
\psi_M \psi_m^\dagger 
\right] \phi
\label{i1}
\ee
%This simplified hamiltonian does not enforce momentum conservation
Since we are dealing with homogeneous fields,
only the energy conservation
is obtained upon looking at the resonance condition, see Appendix A.
%, see below, eq. (\ref{}).

We study the amplitude of spontaneous excitation
which vanishes in a static universe.
% when evaluated in the vacuum of the radiation field $\phi$. 
By this transition, one light massive quantum of mass $m$ 
emits a photon $\phi$
and becomes more massive. Thus if the initial state is
%his amplitude to jump from the state
$\ket{n_i}= \ket{n_M} \ket{n_m} \ket{n_\gamma} $, the final state is
$\ket{n_f}= \ket{n_M + 1 } \ket{n_m-1} \ket{n_\gamma +1}$.
%since one quantum of $\phi$ is added. 
We have not assumed that the fields
 be in their vacuum state. 
%This generalization offers no difficulty.
The motivation for this generalization which offers no
difficulty is the following. 
In quantum cosmology, the number of quanta plays two different roles.
First, as in the BFA, it determines the norm of the amplitudes, 
 see eq. (\ref{me}).
More importantly, when $a$ is treated as a dynamical 
variable, the energy of these 
quanta influences its propagation.
Therefore, in order to exhibit both roles, we shall consider $n_i \neq 0$.

In the {\it interacting } picture, this amplitude is given by
\begin{eqnarray}
{B}^{n_i}(t_2, t_1 ;k) &=&  
%\bra{n_M + 1} \bra{n_m-1} \bra{n_\gamma +1}
\bra{n_f}
e^{-i \int_{t_1}^{t_2} dt H_{int}} 
%\ket{n_M} \ket{n_m} \ket{n_\gamma}
\ket{n_i}
\label{junk}
\ea
where $e^{-i \int_{t_1}^{t_2} dt H_{int}}$ is a time ordered product.
To first order in $g$ one has
\begin{eqnarray}
{B}^{n_i}(t_2, t_1 ;k) &=&  -i
%g \int_{t_1}^{t_2} dt \ \bra{n_f}  \psi_M^\dagger (t) \psi_m  (t) \phi (t) \ket{n_i}
\int_{t_1}^{t_2} dt \ \bra{n_f} H_{int}(t) \ket{n_i}
\nonumber\\
&=& -i \int_{t_1}^{t_2} dt \  e^{-i \int_{t}^{t_2} dt E(n_f , t)} \
\bra{n_f} H_{int} \ket{n_i} \
e^{-i \int_{t_1}^{t} dt E(n_i , t)}
\nonumber\\
%&=& -ig \ \bra{n_f}  \psi_M^\dagger \psi_m  \phi \ket{n_i} \ 
%\int_{t_1}^{t_2} dt \ e^{-im (t - t_1) } e^{-iM (t_2 - t)} e^{-i k \Delta \eta (
%t_2, t)} \ 
%\nonumber\\
&=& -ig \ \bra{n_f}  \psi_M^\dagger \psi_m  \phi \ket{n_i} 
\int_{t_1}^{t_2} dt \ e^{-i \Delta m (t_2 - t)} e^{-i k \Delta \eta (
t_2, t)} \times \left(e^{-i \int_{t_1}^{t_2} dt E(n_i , t)} \right) \quad
%\bra{n_f}  \psi_M^\dagger \psi_m  \phi \ket{n_i}
%\sqrt{ n_m (n_M + 1) (n_\gamma +1) / 8 k}
%\nonumber\\&=&  -ig
%\int_{t_1}^{t_2} dt e^{-i \Delta m (t_2 - t) } e^{-i k \Delta \eta (
%t_2, t)} \sqrt{ n_m (n_M + 1) (n_\gamma +1) / 8 k} \left( e^{-im(t_2 - t_1)}
%\right)
\label{calB}
\end{eqnarray}
where $E(n_i , t) = \bra{n_i} H_0 (t) \ket{n_i}$ is the eigenvalue of
free hamiltonian of the 
state $\ket{n_i}$, see eq. (\ref{blu}),
and $E(n_f , t)$ the energy of the final state.
$\Delta \eta(t_2, t)$ is the lapse of conformal time.
In the third line, we have factorized 
%the phase 
$\exp{-i \int_{t_1}^{t_2} dt E(n_i , t)}$. This phase 
arises from the free evolution of the initial state
from $t_1$ to $t_2$
and  plays no role upon computing the rate of transition. We
shall return to the meaning of this factor upon considering the same 
amplitude in quantum cosmology.
%, in Section 3.
We have also introduced the (Schroedinger) matrix element 
$\bra{n_f}  \psi_M^\dagger \psi_m  \phi \ket{n_i}$ 
evaluated with time independent operators.
%$\bra{n_f}  \psi_M^\dagger \psi_m  \phi \ket{n_i}$. 
It is equal to
\be
\bra{n_f}  \psi_M^\dagger \psi_m  \phi \ket{n_i} =
{  \sqrt{ n_m (n_M + 1) (n_\gamma +1)}
\over \sqrt{2M 2m 2k}}
\label{me}
\ee
We have presented this calculation \` a la Schroedinger
 because, in quantum cosmology, the notion of an
 Heisenberg operator no longer exists,
see the remark 3, Section 4 in \cite{wdwgf}. 
%As an illustration of this absence
For the same reason, 
%we shall see in Section 3 that 
amplitudes of transition
will be delivered in a Schroedinger form 
similar to the one of the second line in
eq. (\ref{calB}) when working in quantum cosmology, in Section 3.

In Appendix A, the amplitude ${B}^{n_i}(t_2, t_1 ;k) $
 is evaluated in the de Sitter case.
%In that background, 
%When the universe is de Sitter dominated, 
%the scale factor $a$ satisfies $\dot a / a = h = \Lambda^{1/2} = constant$ 
%and the lapse of
%conformal time lapse is $\Delta \eta (t_2, t)= - e^{-h t_2} + e^{-h t}$.
%Therefore, the amplitude $B$ is 
%\be
%{B}^{n_i}(t_2, t_1 ;k) = -ig \ \bra{n_f}  \psi_M^\dagger \psi_m  \phi \ket{n_i} \ 
%\int_{t_1}^{t_2} dt \  e^{-i \Delta m (t_2 - t) } 
%e^{-i k (e^{h t_2} - e^{h t})} \times
% \left( e^{- im(t_2 - t_1)} \right)
%\label{calB2}
%\ee
%In the limit $t_1 \to - \infty$, $t_2 \to \infty$, this
%amplitude can be computed analytically, 
%see Appendix A. 
The computation 
%of the amplitude
has been relegated to an Appendix
in order not to deviate from our main goal 
%the discourse 
which is to determine 
the modifications of the amplitudes introduced by 
the dynamical character of gravity.  Therefore,
we only consider schematically the probability 
and the rate of transitions.

The weight of $a(t)$ in $H_{int}$
has been chosen such that the Golden Rule is recovered,
 upon considering a {\it collection} of massless
excitations of density\footnote{
We could have equally work with the $3$-dimensional weight of $a(t)$
 given by the hamiltonian density
$H_{int}^{(3)} = a^3 \tilde \psi_M \tilde \psi^\dagger _m \tilde \phi
= \psi_M  \psi^\dagger _m \phi /a $
where the tilde fields are the unrescaled fields\cite{wdwgf}. 
The amplitudes in that case are related 
to ours by derivative with respect to $k$ since this derivative brings 
a factor of $1/a$.
Thus upon considering the probability,
 the $3$-dimensional density of states $d^3 k = k^2 dk$ brings a 
 factor of $k^2$ which can be absorbed by integration par part with respect to the two 
derivatives arising from the square of the amplitudes. This is how a linear 
increase in $\Delta t $ is also found in that case.} $dk/2 \pi$.
Indeed the probability
to have a transition in the conformal vacuum increases linearly
with $\Delta t$ since
\ba
P_B ( t_2 , t_1) &=& \int_0^\infty {dk \over 2 \pi }
 |B^{\tilde n_i}(t_2, t_1 ;k )|^2
\nonumber\\
&=&  {g^2 n_m ( n_M +1) \over 2M 2 m}
% \ \bra{n_f}  \psi_M^\dagger \psi_m  \phi \ket{n_i} \
\int_{t_1}^{t_2} dt_b  \int_{t_1}^{t_2} dt_a
\  e^{-i \Delta m (t_b - t_a) } \int_0^\infty dk  {
e^{-i k \Delta \eta (t_b, t_a) } \over 4 \pi k}
\nonumber\\
&\simeq& {g^2 n_m ( n_M +1) \over 2M 2 m}
 \int_{t_1}^{t_2} d(t_b + t_a) \int_{-\infty}^\infty d(t_b - t_a)
e^{-i \Delta m (t_b - t_a) }  G_\phi (t_b, t_a)
%\nonumber\\
%&=&
%(t_2 - t_1) g ^2 \int_{-\infty}^\infty d(t_b - t_a)
%e^{-i \Delta m (t_b - t_a) } \ln( e^{- h t_b } - e^{- h t_a})
%e^{-i \Delta m (t_b - t_a) } \ln( e^{- h t_b } - e^{- h t_a})
\nonumber\\
&=& n_m ( n_M +1) (t_2 - t_1)  R
\label{GR1}
\ea
%\int_0^\infty dk =  \int_{k_1}^{k_2} { dk \over k} g ^2 R = (t_2 - t_1) g ^2 R
%where $k_i = \Delta m e^{ h t_i}$ are the minimal and maximal frequencies
%which interfere with $\Delta m $ during the interval $t_2, t_1$ and
where $R$ is the rate of transition for one atom.
 Up to a prefactor, it is
given by the Fourier transform
of the (positive)  Wightman function $G^+_\phi (t_b, t_a) =
 (4 \pi )^{-1} \ln \Delta \eta (t_b, t_a)  $ 
with respected to $e^{-i \Delta m t}$, 
see refs. \cite{unruh}\cite{BD}\cite{GO} for more detail.
In the third line, we have extended the domain of integration 
of $t_b - t_a$ 
to $[-\infty, \infty]$ in conformity to the 
Golden Rule limit.

We now generalize eq. (\ref{GR1}) in the case wherein the 
number of massive quanta is not exactly known. Suppose
that the occupation numbers $n_M, n_m$ have amplitudes
$c_{n_M, n_m}$. Then the probability for having a transition from one $m$ to 
$M$ in the lapse of proper time $\Delta t =t_2 - t_1 $ is simply 
given by the weighted sum of the probabilities
\ba
\bk{P_B ( t_2 , t_1)} &=&
\Sigma_{n_M, n_m} \ \vert c_{n_M, n_m} \vert^2 \ P_B^{n_i} ( t_2 , t_1)
\nonumber\\
&=& ( R \ \Delta t ) \  \Sigma_{n_M, n_m} \ \vert c_{n_M, n_m} \vert^2 \ 
n_m ( n_M +1)
\label{sump}
\ea
The rate $R$ times the lapse of time factorizes in this treatment
wherein the dynamics of gravity is completely neglected.
This factorization will no longer be found upon
taking into account this dynamics at the quantum
level, i.e. by abandoning the BFA. See Section 5 for the proof.

%In section 4, upon taking into account the momenta
%of the quanta, we shall use the usual hamiltonian given 
%in terms of the unrescaled fields by 
%$\int d^3 x \sqrt{ -g} \tilde \psi_M  \tilde \psi_m \tilde \phi$. It 
% differs from eq. (\ref{}) by a factor of $a(t)$.
%However the density of states of the radiation field $\tilde \phi$ is
% $d^3 k  = dk k^2$ in a flat universe. This guarantees
%that the probability $P_A ( t_2 , t_1)$ increases linearly with 
% the proper time  of the detector.
%is recovered.

%\section{Time dependent perturbation theory in Quantum Cosmology}
\section{The amplitudes in Quantum Cosmology}

We shall define and compute the transition amplitudes of the same
model in Quantum Cosmology, by using only
the solutions of the Wheeler De Witt (WDW) equation. 
Our aim is to determine, both conceptually and quantitatively,
 what are the consequences of the enlargement of the 
quantum dynamics to the propagation of $a$, i.e. 
the consequences of
the abandonment of the  
background field approximation. 

To define transition amplitudes in quantum cosmology, we 
include the interaction hamiltonian 
in the WDW constraint. This is mandatory. Indeed reparamatrization
invariance requires that the sum of all forms of energy vanishes.
Then gravity, i.e. the propagation of $a$,
%the cosmological expansion in our simplified model, 
is modified by this additional matter hamiltonian.
%Hence, gravity ``incorporates'' the matter interactions. 
This modification 
%of the propagation of $a$
 is now essential
in order to retrieve matter interactions from the 
{\it correlations} between gravity and matter.
We emphasize this point. Whereas matter interactions
are formulate only in terms of matter operators in the BFA,
see Section 2, upon dealing with a reparamatrization
invariant quantum framework, these interactions are now
represented as {\it quantum correlations} among gravity and matter.

To analyse these correlations and to 
%recuperate this eaten up interaction hamiltonian, we shall perform
recover perturbation theory in a well defined scheme of approximations,
 we shall perform a first order development in the modification of 
the propagation of gravity induced by the interaction hamiltonian. 
This will deliver the transition amplitudes in the framework of
quantum cosmology.

More explicitly, we shall proceed in two steps. 
We shall first use a treatment based on the
hypothesis that the propagation of $a$
can be properly described by a WKB wave, i.e. by a first
order differential equation. This treatment
 is both mathematically and conceptually straightforward.
Furthermore it clearly
illustrates how the modification of the propagation of $a$
delivers time dependent perturbation theory. 
The second treatment is more rigorous and 
based directly on the (second order) WDW constraint
wherein the matter hamiltonian contains the 
interaction term. However it requires some work
to understand how probability amplitudes should be extracted 
from the combined evolution of matter and gravity.
(This is why we shall first present the first order analysis.)
By applying a perturbative
treatment of the solutions of the second order constraint equation, 
we shall show how to 
recover the expressions obtained in the first approach.
%and thereby justify them.
% a posteriori.
Finally,  we shall prove that the amplitudes so extracted coincide with
the expressions evaluated at the BFA when one proceeds to a 
{\it double} development\cite{wdwgf}: 
a WKB approximation for the free propagation of gravity and 
a first order development in the energy change induced by the 
transition itself. The quantitative corrections to the transition amplitudes
are therefore controlled by the corrections to both of these
approximations. 
They are studied in Section 5.

Quantum cosmology is based on the solutions of the (WDW) 
constraint equation 
%The total hamiltonian of matter at $a$ is
\be
 \left[ H_G + H_{matter} \right] \Psi = 0
\label{o}
\ee

In ref. \cite{wdwgf} a simple model based on the free 
%($g= 0$)
hamiltonian, eq. (\ref{blu}),
was extensively studied. In particular Green functions
were evaluated from the solutions of eq. (\ref{o})
for both the massive fields and the massless conformal field.
In what follows, we shall use the same techniques to 
define and compute transition amplitudes.

To obtain these amplitudes, one should consider
the modified WDW equation wherein the 
matter hamiltonian is now the sum of the free, eq. (\ref{blu}),
and the interacting eq. (\ref{i1}), hamiltonians.
Explicitly, the modified WDW equation reads, see eq. (23)
in \cite{wdwgf}, 
\be
\left[
-G^2 \partial_a^2 + \kappa a^2 + \Lambda a^4
   + 2Ga ( H_{0} + H_{int} ) \right] \Psi (a, \psi_M, \psi_m, \phi) = 0
\label{modwdw}
\ee
where the matter hamiltonian operator is
%\left( d_M^\dagger d_M + m d_m^\dagger d_m + { k \over a } 
% d_\gamma^\dagger d_\gamma + g (d_M^\dagger d_m + d_m^\dagger d_M )
%(d_\gamma^\dagger +  d_\gamma ) \right)
\be
 H_{0} + H_{int} = 
M d_M^\dagger d_M + m d_m^\dagger d_m + { k \over a } 
 d_\gamma^\dagger d_\gamma + 
g \left[ \psi_M^\dagger \psi_m+
\psi_M \psi_m^\dagger 
\right] \phi
%g { (d_M^\dagger d_m + d_m^\dagger d_M )
%(d_\gamma^\dagger +  d_\gamma ) \over {\sqrt{ 2M 2 m 2 k}}}
\label{blu'}
\ee
%where we have introduced the destruction and creator operators of
%the three fields and remove the $1/2$ for simplicity.
%see eqs. (\ref{blu}, \ref{i1}). We recall that t
The operator $d_M$ ($d_M^\dagger$) 
should be read as $ ( i \partial_{\psi_M} \pm M  \psi_M )/ \sqrt{2M} $
in the field representation.
%One should be cautious when interpreting the absence 
%of dependence on $a$ in $ H_{int}$. Indeed  $ H_{int}$ does depend on $a$
%through the fields.
% since in the interacting picture. 
%In order to obtain correctly this hidden dependence, 
%we shall be lead to work in the Schroedinger picture
%and use the matrix element eq. (\ref{me}) evaluated at single time, 
%i.e. at a given $a$.
%It will be only upon {\it evaluating} a specific amplitude
%that the precise meaning of this hamiltonian operator will
%be given. We therefore ask the reader to have some patience
%and to view the following developments as a formal book
%keeping.

Our aim is to extract the amplitudes of transitions from the
combined evolution of gravity and matter
described by the solutions $ \Psi (a, \psi_M, \psi_m, \phi) $.
As said above,
we shall proceed in two steps.

\vskip .3 truecm
{\it{The approach based on a first order equation}}

\noindent
This first approach relies on the hypothesis
that the propagation of $a$ is accurately described by 
WKB solutions. In this respect, we recall that one of the main 
results of \cite{wdwgf} was to show that the validity of the WKB 
approximation is ``extended'' in the sense
that only a small part of the 
corrections to this approximation do modify matter matrix elements.
This leads to the important fact that
the modifications of matrix elements due to these corrections
%of the WKB approximation 
are irrelevant in the case of 
macroscopic matter dominated universes.
%The same holds in the present case.

% n the limit $g \to 0$, one recovers the model of \cite{wdwgf} of three 
%harmonic oscillators coupled to gravity through the WDW constraint only.
In a WKB treatment, 
%for the propagation of $a$ 
eq. (\ref{modwdw}) 
determines the momentum operator
%(we have in mind a WKB treatment for the propagation of $a$ only) is
\ba
\hat \pi(a , g) &=& - G^{-1} \sqrt{ \kappa a^2 + \Lambda a^4 
+ 2 a G \left( H^{0}_{} + H_{int} 
\right)}
\label{pii'}
\ea
Since $\hat \pi(a , g)$ incorporates $H_{int}$, the propagation of $a$ is modified
by this hamiltonian at the quantum level. 
In this, we see that gravity is sensitive
to the hamiltonian operator and not only to
its expectation value like in the semi-classical approximation. 
To evaluate this q-number dependence, 
we develop $\hat \pi$ in powers
of $H_{int}$:
\ba
\hat \pi(a , g) &=& \pi(a, \hat n_j) + H_{int} { a  \over G \pi(a, \hat n_j) }
 -  H^2_{int} {  a^2  \over G^2 \pi(a, \hat n_j)^3 }
%{(d_M^\dagger d_m + d_m^\dagger d_M )^2
%(d_\gamma^\dagger +  d_\gamma )^2 \over  2M 2 m 2 k} 
+ {\cal{O}}(g^3)
\label{pii}
\ea
where $\hat n_j$ designates the counting operators of the
three fields.

The simplest way to describe the combined propagation of
gravity and matter, is to construct the evolution operator ${\cal{U}}(a_2, a_1)$.
In the classification presented by Isham \cite{isham}, this approach 
corresponds to ``identify time before quantisation''. 
This means that the operator ${\cal{U}}(a_2, a_1)$ satisfies
 the first order equation 
\be
i \partial_a \ {\cal{U}}(a, a_1) = \hat \pi(a , g) \ {\cal{U}}(a, a_1) 
\label{firsto}
\ee
with the usual boundary condition ${\cal{U}}(a, a) = 1$.

In terms of the free ($g=0$) evolution operator ${\cal{U}}^{0}(a, a_1)$,
defined by
\be
i \partial_a \ {\cal{U}}^{0}(a, a_1) = \pi(a, \hat n_j) \ {\cal{U}}^{0}(a, a_1)
\label{firstofree}
\ee
%which is diagonal in the Fock space of matter,
we can write ${\cal{U}}(a_2, a_1)$ as 
\be 
{\cal{U}}(a_2, a_1) = {\cal{U}}^{0}(a_2, a_1) - i
\int_{a_1}^{a2} da \ {\cal{U}}^{0}(a_2, a)  \left( 
H_{int} { a  \over G \pi(a, \hat n_j) } \right) 
{\cal{U}}^{0}(a, a_1) + {\cal{O}}(g^2)
\label{udevel}
\ee
%Eq. (\ref{stuf}) 
This is in the strict analogy with the 
%perturbative development
%of the 
evolution operator in quantum mechanics evaluated 
perturbatively in the
Schroedinger picture, see eq (\ref{calB}). 
Indeed, when the
total hamiltonian is decomposed as $H_0 + H_{int}$, 
the full evolution operator $U$
is given in terms of the unperturbed operator $U_0$ as
%, to first order in $\delta H$ by
\ba
U(t_2, t_1) = U_0(t_2, t_1) - i \int_{t_1}^{t_2} dt \ U_0(t_2, t) \ H_{int} \ 
U_0(t, t_1) + {\cal{O}}(g^2)
%\nonumber\\ && \quad - 
%{ 1 \over 2}
%\int_{t_1}^{t_2} dt_a \int_{t_1}^{t_a} dt_b \ 
%U_0(t_2, t_a) \ H_{int} \  U_0(t_a, t_b) \ H_{int} \ U_0(t_b, t_1)
\label{UU}
\ea
Thus the amplitude 
${\cal{B}}^{n_i}_{first}(a_2, a_1 ; k)$ 
 to start from $a= a_1$ with the matter content specified by
$\ket{n_i}$ and to end up at 
$a= a_2$ with $\ket{n_f}$ is given by
by the matrix element of
${\cal{U}}(a_2, a_1)$. 
In this {\it first} order formalism with respect to $i \partial_a$
one finds
\ba
{\cal{B}}^{n_i}_{first}(a_2, a_1 ; k) &=&
-i \int_{a_1}^{a_2} \! da \
 {\cal{U}}^{0}(a_2, a ; n_f ) \
\bra{n_f}   {  a H_{int}
 \over G \pi(a, \hat n_j)} \ket{n_i} \  {\cal{U}}^{0}(a, a_1; n_i)
\label{firsto2}
\ea
where we have used the diagonal character of ${\cal{U}}^{0}(a', a)$
and defined
\ba
{\cal{U}}^{0}(a, a_1; n_i) = \bra{n_i} \ {\cal{U}}^{0}(a, a_1) \ket{n_i}
%\nonumber\\
= e^{ i \int_{a_1}^{a}
\pi(a', n_i)  da'   }
\label{first3}
\ea
The classical momentum $\pi(a, n_i)$ is the solution of eq. (\ref{pii'}) at $g=0$
in the initial state characterized dy the occupation numbers $n_i$,
i.e. $\pi(a, n_i) = \bra{n_i} \pi(a, \hat n_j) \ket{n_i}$.
We now define the ambiguous expression $\bra{n_f}{ H_{int}
 / \pi(a, \hat n_j)} \ket{n_i}$ by the following expression
\be
{\bra{n_f}{ H_{int} \over \pi(a, \hat n_j)} \ket{n_i}} = 
{ \bra{n_f} H_{int} \ket{n_i} \over \sqrt{ \pi(a, n_i) \pi(a, n_f) }}
\label{amb}
\ee
where the matrix element of $H_{int}$ is given in eq. (\ref{me}).
We shall justify our 
choice in the more rigorous second approach. This type of ambiguity
is inherent to the present approach wherein the full hamiltonian
is given by square root, eq. (\ref{pii'}), see \cite{recR} for a similar treatment
based on a first order equation applied to a relativistic oscillator.

Using these expressions, ${\cal{B}}^{n_i}_{first}(a_2, a_1 ; k) $ becomes
\ba 
{\cal{B}}^{n_i}_{first}=
-i \int_{a_1}^{a_2} \! { da \ (a/G) \over 
\sqrt{ \pi(a, n_i) \pi(a, n_f) }} \
e^{ i \int_{a}^{a_2}
\pi(a', n_f)  da'   } \ \bra{n_f} H_{int}  \ket{n_i} \ 
e^{ i \int_{a_1}^{a}
\pi(a', n_i)  da'   } 
%\bra{n_f} H_{int}  \ket{n_i}
%\int_{a_1}^{a_2} \! { da  (a/G) \over 
%\sqrt{ \pi(a, n_i) \pi(a, n_f) }} e^{ i \int_{a}^{a_2} \left[
%\pi(a', n_f) -  \pi(a', n_i)  \right] da'   } 
%\times e^{ i \int_{a_1}^{a_2}
%\pi(a, n_i)  da  }
\label{firsto22}
\ea
The comparison of eqs. (\ref{firsto2}, \ref{firsto22})
with the expression evaluated at the 
BFA, eq. (\ref{calB}), is very instructive. What is not modified by the
dynamical character of $a$ is the
structure of the expression. The amplitude is given of the
integral over the parameter $a$ (in the place of $t$) of three factors:
the two free kernels transporting the initial and the final matter
configurations from the end points to the ``place'' where
the intercation occurs and the matrix element  of $H_{int}$
evaluated at that place, in the Schroedinger picture.
What is modified, is the phase of these free propagations and the
norm weighting the integration over $a$.
%  we see that the changes are the following. The m 

We postpone the evaluation of eq. (\ref{firsto22}).
%this expression and the comparison with eq. (\ref{}). 
We shall
first rederive it with more rigor, without making
appeal to the first order (WKB) hypothesis.
This shall afford us new insights about the nature
of quantum gravity.

\vskip .3 truecm
{\it{ The approach based on the second order equation}}

\noindent
Our aim is to deduce transition amplitudes
from the combined evolution
of gravity and matter described by the
second order WDW equation, eq. (\ref{modwdw}). 
To this end
%, as in \cite{wdwgf}, 
it is convenient to introduce
the ``kernel-operator'' ${\cal{K}}(a_2, a_1)$,  solution
of
\be
\left[
-G^2 \partial_a^2 + \kappa a^2 + \Lambda a^4
   + 2Ga ( H_{0} + H_{int} ) \right] {\cal{K}}(a_2, a_1) = i G^2
\delta ( a_2 - a_1)
\label{second1}
\ee
with unit wronskian in the following sense
\be
\lim_{a_2 - a_1 \to 0^+}
\left( \partial_{a_2} - \partial_{a_1} \right)
{\cal{K}}(a_2, a_1 ) = - i 
\label{M17p}
\ee
The choice of the factors of $i$ in the r.h.s. of 
eqs. (\ref{second1}, \ref{M17p})
is purely conventional. More important is that these equations do not
fixed univocally the kernel ${\cal{K}}(a_2, a_1 )$. To 
properly identify what is the
additional requirement, we refer to the treatment of the free ($g=0$)
kernel presented in \cite{wdwgf}. In that case, the boundary
condition was obtained from the requirement
that the constraint equation be imposed by integrating 
over {\it positive} constant lapse $N$ only, see eq. (27) therein. 
This implies that,  in the semi classical regime,
for $a_1 < a_2$, only one dominant
contribution is obtained, the one
associated with an expanding universe. Had we integrated over all $N$
we would have obtained the contributions of both the 
expanding and contracting motion.
Therefore, the restriction to positive lapse $N$ introduces
the notion of ``ordered propagation'' in the sense that it selects
one runing wave instead of a stationary (real) sine or cosine.
This is of crucial importance for what follows the next paragraph.

We recall that free propagation is described 
by a diagonal kernel. Thus when the matter state is characterized
by  $n_i$ quanta, the matrix element of the
free kernel is given by a c-number function which reads in the
WKB approximation,
%analysed in \cite{wdwgf} and given, 
%in the WKB approximation, it reads
%is given by the c-number function
\ba
{\cal{K}}^0(a_2, a_1; n_i) =   \bra{n_i} {\cal{K}}^0(a_2, a_1) \ket{n_i} =
%\tilde Tr_{g=0} \left[ \ket{a_1} 
%\ket{n_M} \ket{n_m} \ket{n_\gamma} \ket{n_i} \ \bra{n_i} \bra{a_2} \ \right] 
 { e^{ i \int_{a_1}^{a_2}
\pi(a, n_i)  da } \over 2 \sqrt{ \pi(a_2, n_i)
\pi (a_1, n_i) }}
%\nonumber\\
%&=& { 1 
= {{\cal{U}}^{0}(a_2, a_1; n_i)
\over 2 \sqrt{ \pi(a_2, n_i)
\pi (a_1, n_i) }}
\label{M16'}
\ea
In the last equality, we have compared the WKB expression
of ${\cal{K}}^0(a_2, a_1; n_i) $ characterized by a unit conserved 
wronskian to the first order operator ${\cal{U}}^{0}(a_2, a_1; n_i)$ possessing 
a unit norm. The extra factors of $1/ \sqrt{  2 \vert \pi(a, n_i)
\vert}$ at both
ends of the propagation 
%define the weights
%attributed to ``to be at $a_i$'' and ``to be at $a_f$''. They 
appear inevitably in this framework since ${\cal{K}}^0(a_2, a_1)$ is a
solution of a second order differential equation of the continuous variable
$a$. 
They will therefore reappear in the sequel and must be
removed when one wants to obtain {\it probability} amplitudes.
% of {\it probability} concerning matter transition. 
This necessary amputation of kinetic factors 
%of $1/ \sqrt{ 2 \vert \pi(a, n_i)\vert}$ at the end points 
%in order to obtain probability amplitudes 
is the ``reduction formula'' applied in the present case.
It is appropriate to remember that
this removal should be performed at the end points only.
Indeed, as we shall see, the solutions of eq. (\ref{second1}) deliver
the correct weight at the intermediate points where the 
interaction hamiltonian acts. Notice also that this amputation 
would be automatically carried out by adopting a ``third''
quantized framework wherein single universe states are 
normalized to one. 

Having specified univocally the free kernel one can express,
as in eqs. (\ref{udevel}, \ref{UU}), the full kernel
${\cal{K}}(a_2, a_1)$ in terms of the free one as
\be
{\cal{K}}(a_2, a_1) = {\cal{K}}^0 (a_2, a_1) - i 
\int_{a_1}^{a_2} \! da \  {\cal{K}}^0(a_2, a) \ { 2a \over G} H_{int} \ {\cal{K}}^0
 (a, a_1) + {\cal{O}}(g^2)
\label{second3}
\ee
The factor of $-i$ in the second term is not an artifact of the
choices adopted in eqs. (\ref{second1}, \ref{M17p}) in the following sense.
Had we redefined ${\cal{K}}^{new}(a_2, a_1)= i {\cal{K}}(a_2, a_1) $ so as to 
obtain two real equations in the place of eqs. (\ref{second1}, \ref{M17p}),
then,  the value of the kernel would have been 
purely imaginary for $a_2 \to a_1$ instead of purely real as in eq. (\ref{M16'}).
Thus the {\it relative} factor between free propagation and the
first order term in $g$ is always purely imaginary, as it should be, 
when considering a perturbation diagonal in $n_i$, i.e. a level shift.
% treated perturbatively.
This is a direct consequence of the above mentioned choice of
integration over positive lapse only. 
Therefore, 
%in the case of interactions 
this choice guarantees
the obtention of the usual Born series, at least to first order in $g$.
%from a perturbative treatment
%of a real equation, eq. (\ref{}), and a real normalization, eq . (\ref{}), 
%looks rather odd. However it is both correct and crucial.
%has the deep origin that the free kernel is 
%We are now in position to define the 

The (un-amputated) amplitude
of transition, ${\cal{B}}^{n_i}(a_2, a_1 ; k)$,
to go from the state $n_i$ at $a_1$ to the state $n_f$ at $a_2$
is given by the matrix element of ${\cal{K}}(a_2, a_1)$
\ba
{\cal{B}}^{n_i}(a_2, a_1 ; k) &=& \bra{n_f} \ {\cal{K}}(a_2, a_1) \ket{n_i}
\nonumber\\  
 &=&
- i \ \int_{a_1}^{a_2} da {2 a \over G } \ {\cal{K}}^0(a_2, a; n_f)
 \ \bra{n_f} H_{int} \ket{n_i} \ 
{\cal{K}}^0(a, a_1; n_i) 
\label{defi}
\ea
Thus, to first order in $g$, but without any other approximation\footnote{
This is not exactly true. Indeed, since the integration over positive lapses
leads to ${\cal{K}}^0(a, a_2)={\cal{K}}^0(a_2, a)$,
% \neq 0$ for $a_2 > a$,
 we should include $Z$ graphs, i.e.  
%in the last two equations,
the contributions from $a$'s  bigger than $a_2$ or smaller than $a_1$.
Their truncation is however legitimate since these graphs 
concern the whole universe and not only a particle as
in QFT. Notice that these ``usual'' $Z$ graphs are present in the
higher terms of eq. (\ref{second3}).}, this
amplitude  
%${\cal{B}}^{n_i}(a_2, a_1 ; k)$ 
%kernel ${\cal{K}}(a_2, a_1)$
decomposes itself
into the convolution of two free kernels and the matrix element
of the interaction hamiltonian 
operator, as in eqs. (\ref{calB}, \ref{firsto2}).
%Schroedinger matrix element of the interaction hamiltonian. 
This is our first result. 
This arrangement determines all the properties of 
%the transitions amplitudes
 ${\cal{B}}^{n_i}(a_2, a_1 ; k)$ 
that we now analyse.

%To further explore the meaning of the amplitude
To this end, 
we shall perform, in what follows,
 a {\it double} development\cite{wdwgf}
which consists in
(i) using WKB expressions for both ``free'' kernels ${\cal{K}}^0$
%, see eq. (\ref{M16}), 
and
(ii) making a 
first order development in the change of the occupation numbers
$n_j$.
The first step will make contact with the amplitude 
${\cal{B}}_{first}^{n_i}(a_2, a_1 ; k)$ 
evaluated in the first order formalism, see eq. (\ref{firsto22}), 
and both steps with 
the amplitude $B^{n_i}(t_2, t_1 ; k)$ evaluated at the BFA , see
%and given in 
eq. (\ref{calB}).
In Appendix B, we present an alternative evaluation of 
${\cal{B}}^{n_i}(a_2, a_1 ; k)$ which does not rely on step (ii) and
which never uses the background concept.

Using the WKB approximation for the free kernel, see
eq. (\ref{second3}), one obtains
\ba
{\cal{B}}^{n_i}(a_2, a_1 ; k) &=& 
%- i \ \int_{a_1}^{a_2} 
% { - i \ \bra{n_f} H_{int} \ket{n_i} 
% \over { 2 \sqrt{ \pi(a_2, n_f) \pi(a_1, n_i)}}}
%\int_{a_1}^{a_2} { da  (a /G) \over \sqrt{ \pi(a, n_f) \pi(a, n_i)}} 
%e^{i \int_a^{a_2} \pi(a', n_f) da'} 
%e^{i \int_{a_1}^a \pi(a', n_i) da'} 
%\nonumber\\  {- i\ \bra{n_f} H_{int} \ket{n_i} 
%\over { 2 \sqrt{ \pi(a_2, n_f) \pi(a_1, n_i)}}} \nonumber\\  &&
- i\ \bra{n_f} H_{int} \ket{n_i} 
% \over { 2 \sqrt{ \pi(a_2, n_f) \pi(a_1, n_i)}}} \nonumber\\  &&\quad 
%\int_{t_1}^{t_2} dt 
\int_{a_1}^{a_2} {a da \over G \sqrt{ \pi(a, n_f) \pi(a, n_i)}}
\ e^{i \int_a^{a_2} [\pi(a', n_f) - \pi(a', n_i) ] da'}
%\times e^{i \int_{a_1}^{a_2} \pi(a', n_i) da'}  
\nonumber\\  
&&
\quad \quad \quad \quad \quad \quad  
\times {1  \over { 2 \sqrt{ \pi(a_2, n_f) \pi(a_1, n_i)}}} \times
e^{i \int_{a_1}^{a_2} \pi(a', n_i) da'}  
\nonumber\\  
&=& { 1 \over 2 \sqrt{ \pi(a_2, n_f) \pi(a_1, n_i)}} \
{\cal{B}}_{first}^{n_i}(a_2, a_1 ; k) 
%\sqrt{{ \pi(a_2, n_i) \over \pi(a_2, n_f) }}
%\times
% \left( {e^{i \int_{a_1}^{a_2} \pi(a', n_i) da'}  
%\over { 2 \sqrt{ \pi(a_2, n_i) \pi(a_1, n_i)}}}
%\right)
%\nonumber\\  
%&=& { 1 \over 2 \sqrt{ \pi(a_2, n_f) \pi(a_1, n_i)}} B^{n_i}(t_2, t_1 ; k) 
%{\mbox{ times a phase}}
\label{better}
\ea
In the first equality, we have factorized both the kinematic
factor that should be amputated and, as in
eq. (\ref{calB}), the global $a$-independent phase
corresponding to the free propagation of the initial state.
In the second equality, we see
that the {\it amputated} amplitude ${\cal{B}}_{amp}^{n_i}(a_2, a_1 ; k)$
given by
 $ 2 \sqrt{ \pi(a_2, n_f) \pi(a_1, n_i)}
{\cal{B}}^{n_i}(a_2, a_1 ; k) $ is equal to the amplitude
evaluated in the first order treatment. 
It is now appropriate to notice that the 
development in eq. (\ref{second3}) is directly expressed in terms of 
the hamiltonian $H_{int}$ and not in terms of the
clumsy operator $H_{int} / \pi(a, \hat n_j)$ as in 
%the first order development, see 
eq. (\ref{udevel}). Therefore, in the
second order formalism, we do not meet the 
ambiguity in defining its matrix elements, c.f. eq. (\ref{amb}).
Indeed, eq. (\ref{defi}) is univocally defined.

We now perform the second step of the double development, that is,
 a first order expansion in the occupation
numbers change characterizing the transition under examination. 
%$\Delta m$ and $k$.
To this order, one can use 
the Hamilton-Jacobi formalism and writes $a$-dependent phase 
in ${\cal{B}}^{n_i}(a_2, a_1 ; k)$ as
%amplitude as
%out because, to first order in $\Delta m$ and in $k$ one has 
\ba
&&\int_{a}^{a_2} \left[ \pi(a', n_M + 1, n_m - 1, n_\gamma + 1) -
\pi(a', n_M, n_m, n_\gamma)
\right] da'
\nonumber\\
&&\quad \quad \quad \quad \quad \quad \quad \quad 
\quad \quad \quad \quad =  [ - \partial_{n_m} +
  \partial_{n_M} +  \partial_{n_\gamma} ]
 \int_{a}^{a_2} \pi(a', n_M, n_m, n_\gamma) da'
\nonumber\\
%&&\quad \quad \quad \quad \quad \quad \quad \quad 
%\quad \quad \quad \quad  =   (M- m) \int_{a}^{a_2}  { da' \ a' \over G \pi(a', n_i) }
%+  k  \int_{a}^{a_2}  {  da' \over G \pi(a', n_i) }
%\nonumber\\
&&\quad \quad \quad \quad \quad \quad \quad \quad 
\quad \quad \quad \quad = - (M- m) \ 
\Delta t (a_2, a ; n_i) -  k \ 
\Delta \eta (a_2, a ; n_i)
\label{difftata}
\ea
{\it by definition} of the proper time and conformal 
time lapses, see eq. (17, 18)
%(\ref{}\ref{}) 
in \cite{wdwgf}.
%We notice that both lapses are evaluated in the universe 
%containing $n_i$ quanta and that this expression is gauge independent.
By applying the same development to the norm of the integrand of 
${{\cal{B}}^{n_i}(a_2, a_1 ; k)}$, the amputated  amplitude can be written
%${\cal{B}}^{n_i}$
\ba
{{\cal{B}}_{amp}^{n_i}(a_2, a_1 ; k)}
\times
 e^{-i\int_{a_1}^{a_2} da \pi(a, n_i )}
%\over (2 \sqrt{ \pi(a_2, n_f) \pi(a_1, n_i)})^{-1}}
 &=& {- i \bra{n_f} H_{int} \ket{n_i} } \int_{t_1}^{t_2} \!dt \
e^{- i \Delta m \Delta t(a_2, a; n_i) } e^{- i  k 
\Delta \eta (a_2, a ; n_i)} \ 
%\times e^{i\int_{a_1}^{a_2} \pi(a_2, a_1, n_i )}
% \left( {e^{i \int_{a_1}^{a_2} \pi(a', n_i) da'}  
%\over { 2  \sqrt{   \pi(a_2, n_i) \pi(a_1, n_i)}}}
%\right)
\nonumber\\
&=&  B^{n_i}( t(a_2), t(a_1); k) \times 
 e^{i \int_{t_1}^{t_2} dt E(n_i, t) } 
% \times  
%{e^{i \int_{a_1}^{a_2} \pi(a', n_i) da'}  
%\over { 2 \sqrt{ \pi(a_2, n_i) \pi(a_1, n_i)}}}\right)
\label{better2}
\ea
since $dt = a da /G \pi(a, n_i)$, see eqs. (\ref{pii'}, \ref{difftata}).
The amputated amplitude differs from $B$ {\it only} 
by the two phase factors describing free evolution.
We emphasize this: By eliminating that part of the propagation which concerns
the free propagation of the initial state, the amputated amplitude
evaluated in quantum cosmology {\it equals} in phase and in amplitude
 the corresponding amplitude evaluated at the BFA
upon performing the double development described before eq. (\ref{better}).
This is our second result.

We conclude this Section by two remarks.

\noindent
Both lapses of time in eq. (\ref{difftata}) are evaluated, {\it
by construction},  in the
universe containing $n_i$ quanta since they are conjugated to the
matter energy changes induced by the transition itself, i.e. $n_i \to n_f$.
Time appears because there is a transition. 
The validity of making a first order expansion in $n_i - n_f$
requires that the universe be macroscopic, see Section 5 for
the proof.

We did not assume that the wave function
was a ``quasi-classical''\cite{hartle}\cite{BV}\cite{ortiz} state, 
i.e. a well defined superposition
of WKB solutions which describes a {\it single} classical universe.
Indeed, we have considered
%, exactly as we did for 
%an accelerated atom in \cite{rec}, 
a completely {\it delocalized}
wave function, since there was no spread in the initial energy.
Therefore, the emergence of a well defined time parametrization
does not required to have the ``wave function'' of the universe
peaked along a classical path. Furthermore, 
since {\it each} component does deliver its {\it own} time parametrization
one can also consider widely spread repartitions of energy.
The same is true
%fact that each component defines a time parametrization is also found 
when one considers the transitions of a two 
level relativistic atom. In that case, it is the proper time of the atom
which emerges in transition amplitudes, see \cite{rec} and Section 4. 
%emergence of a time parametrization 
See also Section 5 for more comments on these decoherence effects.
%Therefore,
%we can say that  the emergence of a time parametrization for 
%transition amplitudes ``decoheres'', see Section 5.

%In that Section, the corrections to the various approximations
%used to obtain eq. (\ref{better2}) are evaluated 
%%with respect to the corresponding transitions amplitudes 
%%evaluated at the BFA, that is the corrections to eq. (\ref{ratioss}),
%%are considered in Section 5. We shall evaluate them
%quantitatively. Then we discuss the peculiar properties
% of quantum gravity revealed these corrections.
%%is  independent of the specification of the initial matter state.

\section{The emergence of spatial displacements from recoils}

In this section, we consider an extension of these
transition amplitudes in which the momenta of the exchanged particles
no longer vanish. The motivation is three-fold. First, it is instructive to 
see how one can combine, in a approximative but coherent 
way, quanta
carrying non-vanishing momenta
together with homogeneous geometries.
Secondly, this extension shall allow us to recover the notion of 
spatial displacement from purely {\it homogeneous} solutions, in a manner
similar to the recovery of time lapse from {\it stationary} solutions,
see eq. (\ref{difftata}).
Thirdly, we shall make contact between
the recoil of an heavy atom induced by some momentum transfer\cite{rec}
and the ``recoil'' of gravity induced by some matter transition.
(The reader not interested by these aspects might read directly
Section 5).
 
To these ends, we use the formalism presented in Appendix A 
of ref. \cite{wdwgf}.
It consists in keeping the homogeneity of gravity 
and by considering the dispersion relation of matter quanta. 
The energy of a quantum of mass $M$ and 
(conserved) momentum $p$ is given at radius $a$
by
\be
\Om (M, p, a) = \sqrt{M^2 + p^2 /a^2}
\label{disp}
\ee
In quantum cosmology,
% upon dealing with homogeneous solutions of the WDW equation, 
the conserved momentum $p$ can only be viewed as a quantum number
characterizing states in Fock space.
Indeed, one needs to treat gravity at background 
field approximation in order to be able to {\it use}
% the space-time equation
%velocity of the particle space-time 
%relation
\be
 \partial_p \Om (M, p, a(t)) = { p \over a^2(t) \ \Om (M, p, a(t))} =
{ d z \over dt } = v_z
\label{velo}
\ee
since the definition of velocity requires, as a obvious preliminary, to
have the notion of position. However, in quantum cosmology 
 there is no {\it a priori} notion of position, since 
 no {\it external} device can introduce it. Indeed upon 
using an external device, i.e. by neglecting 
 its momentum content when computing the solutions of the WDW equation,
one would effectively leave the space of gauge invariant solutions. 
%of the WDW equation 
%and use its momentum as a external source.
% whose role is precisely to give a meaning to position. 
It is the necessity of considering all 
devices as {\it internal} constituents
which characterizes quantum cosmology based on the WDW equation.  
% would be 
%inconsistent to neglect its momentum 
Therefore, distances and velocities should be recovered from
internal phase correlations, in a manner similar to the recovery of time.
%the interpretation of the momentum of the particle.

%The validity of the approximative scheme consisting in neglecting
%completely the non uniform part of the 
%%Wheeler De Witt 
%four momentum constraint
%relies on the fact that when working around homogeneous three surfaces, the
%non uniform part of the lapse and the shift functions are first order while
%the zero-momentum part of the lapse is zeroth-order, i.e does not 
%vanish in a completely homogeneous geometry.
%This asymmetry is what legitimizes the present treatment.

The new aspect with respect to Appendix A 
of \cite{wdwgf} concerns the interaction hamiltonian.
%n the Schroedinger picture it 
It is given by
\ba 
\tilde H_{int} = g  \int d^3 p \int d^3 k
 { \left( d_{M, p} d^\dagger_{m, - p- k} 
+ d^\dagger_{M, p}  d_{m, - p- k} \right)
 \left(
d_{k} + d^\dagger_{-k} \right) \over 
\sqrt{ 8\ \Om (M, p, a)\ \Om (m, p+k , a)\ |k| }}
\label{d1}
\ea
where $d_{M, p}$ destroys a massive quantum of momentum $p$ and mass $M$,
$d^\dagger_{m, - p- k} $ creates a massive quantum of
 momentum $- p- k$ and mass $m$,
and $d_{k}$ creates a conformal massless quantum of momentum $k$.
%Notice that we are still dealing with rescaled massive fields related to
%the original $3$-dimensional fields $\tilde \psi$ by 
%$\psi = a^{3/2} \tilde \psi$.
%The extra factor of $a$ comes from the fact that 

%The new important aspect is that 
$\tilde H_{int} $ carries zero momentum.
This is viewed, in the BFA, as a consequence
of the homogeneity of the background. 
%When using the dynamical character of gravity, a new 
In quantum cosmology, a new 
explanation which no longer refers to the notion of a background is required.
%This this vanishing has a more fundamental interpretation.
It is simply that $\tilde H_{int} $ must commute with the total $3$-momentum 
operator which annihilates physical states.
Then, by applying $\tilde H_{int}$ on a zero momentum state, i.e. a 
solution of the WDW equation, the new state remains a 
solution.
% with zero momentum. 
We believe that this new
interpretation based on gauge invariance 
%which no longer refers to a background,
 remains valid when considering 
%non-homogeneous geometries, i.e. 
solutions of the WDW equation wherein
gravitational configurations carry momentum. 
%In that case, the gravitational momentum must be included
%in the extended conservation law.

This commutation allows us to proceed as in Section 4.
The total matter hamiltonian is now
\ba
\tilde H_{matter} &=& H^{0} + \tilde H_{int} = \int d^3 p \ 
\Om( M, p , a) \  d_{M, p}^\dagger d_{M, p} + \int d^3 p \ 
 \Om( m , p , a)\  d_{m, p}^\dagger d_{m, p} 
\nonumber\\ &&
\quad \quad\quad \quad\quad \quad  + \int d^3 k \ 
{ |k| \over a }\ 
 d_{\gamma, k}^\dagger d_{\gamma , k} +
\tilde H_{int} 
\label{blu'1}
\ea
in the place of eq. (\ref{blu'}), and we are interested by the 
amplitude ${\cal{B}}^{n_i}(a_2, a_1 ; p , k)$ in which one massive 
quantum of momentum $p$ emits a photon of momentum $k$. 
This amplitudes connects the initial state
\be
\ket{n_i} = \ket{ n(M, p- k ) , n(m, p), n(\gamma, k) ; \tilde n_i  }
\label{ini}
\ee
defined at $a=a_1$ to the final state
\be
\ket{n_f} = \ket{ n(M, p- k ) + 1, n(m, p) - 1 , n(\gamma, k) + 1 ; \tilde n_i   }
\label{final}
\ee
defined at $a=a_2$. We designate by $\tilde n_i$ the set of spectator quanta
not involved in the transition.
As in eq. (\ref{defi}), the amplitude is defined by
\ba
{\cal{B}}^{n_i}(a_2, a_1 ; p, k)
%\bra{n_M + 1} \bra{n_m-1} \bra{n_\gamma +1} \bra{a_2}
%{\cal{\hat K}} \ket{a_1} \ket{n_M} \ket{n_m} \ket{n_\gamma}
%\nonumber\\
 =   \bra{n_f} \ {\cal{K}}(a_2, a_1) 
%\ket{n_M} \ket{n_m} \ket{n_\gamma}
\ket{n_i} 
% \bra{n_f} \bra{a_2} \ \right]
%\bra{n_M + 1}  \bra{n_m-1} \bra{n_\gamma +1} \bra{a_2} \right]
\label{formal1}
\ea
To first order in $g$, using the WKB expression for
the free kernels, see eq. (\ref{M16'}), the {\it amputated} amplitude
${\cal{B}}_{amp}^{n_i}(a_2, a_1 ; p, k)$
is given by
%one obtains as before
\be
{\cal{B}}_{amp}^{n_i}(a_2, a_1 ; p, k) = -i 
% \sqrt{{ \pi(a_2, n_i) \over \pi(a_2, n_f) }}
\int_{a_1}^{a_2} {  a da 
%\bra{n_f} \tilde H_{int}(a) \ket{n_i}
 \over G \sqrt{ \pi(a, n_f) \pi(a, n_i)}}\ 
\bra{n_f} \tilde H_{int}(a) \ket{n_i}\
e^{i \int_a^{a_2} [\pi(a', n_f) - \pi(a', n_i) ] da'}
%\sqrt{{ \pi(a_2, n_i) \over \pi(a_2, n_f) }}
% \nonumber\\
%&&\quad \quad \quad \quad \quad \quad  \quad \quad \quad \quad 
%\sqrt{{ \pi(a_2, n_i) \over \pi(a_2, n_f) }}
%\times
% \left( {e^{i \int_{a_1}^{a_2} \pi(a', n_i) da'}  \over { 2 \sqrt{ \pi(a_2, n_i)
% \pi(a_1, n_i)}}}
%\right)
%\nonumber\\
%&=& { 1 \over 2 \sqrt{ \pi(a_2, n_f) \pi(a_1, n_i)}} B^{n_i}(t_2, t_1 ; k)
%{\mbox{ times a phase}}
\label{better1}
\ee
see eq. (\ref{better}).
 We have not written the irrelevant phase describing the
free propagation of the initial state.
In the present case
 $\bra{n_f} \tilde H_{int} (a) \ket{n_i}$ is $a$ dependent and given by
\be
\bra{n_f} \tilde H_{int}(a) \ket{n_i} = g \ 
{ \sqrt{ (n(M, p- k ) + 1) \  n(m, p) \
 (n(\gamma, k) + 1)}
  \over 
\sqrt{ 8 \ \Om (M, p-k, a)\ \Om (m, p , a)\ |k| }} 
\label{adep}
\ee
This dependence in $a$ will play a crucial role in recovering the
Lorenz contraction factor, see below.
%As in eq. (\ref{better}) we have factorized out the free propagation of the 
%initial state.

We proceed as in eq. (\ref{difftata}) to a first order development in the 
 energy changes appearing in the
%present in the 
phase of the amplitude.
One obtains
\ba
\int_{a}^{a_2} \left[ \pi(a', n_f) - \pi(a', n_i) \right] da'
%\nonumber\\
%&&\int_{a}^{a_2} \left[ \pi(a', n(M, p- k ) + 1, n(m, p) - 1 , n(\gamma, k) + 1 
%,  \tilde n_i  ) -
%\pi(a', n(M, p- k ) ,  n(m, p) ,  n(\gamma, k),  \tilde n_i  ) 
%\right] da'
%\nonumber\\
%&&\quad \quad \quad \quad
%\quad \quad \quad \quad 
&=&  \int_{a}^{a_2}  { da' \ a' \over G \pi(a', n_i) }
\left[ \Om( M, p -k , a' ) - \Om(m, p, a')  - {|k| \over a' }\right]
\nonumber\\
&=& 
\int_t^{t_2} dt' \left[ \Om( M, p -k , a(t') ) - \Om(m, p, a(t') )  - 
{|k| \over a(t')  }\right]
\quad \
\label{diffta1}
\ea
%where we have introduced the proper time defined, in the universe characterized
%by $n_i$ quanta, by $dt =a da / G \pi(a, n_i)$, see eq. (\ref{difftata}). 
By the same change of variable from $a$ to $t(a)$,
 the norm 
%$a da / G \pi(a, n_i) = dt$ 
in the integration of eq. (\ref{better1}) becomes unity.
%: $a da / G \pi(a, n_i) = dt$.
Thus,
 the amputated amplitude ${\cal{B}}^{n_i}(a_2, a_1 ; p, k) $
 is equal to the corresponding amplitude evaluated at the BFA
in the background characterized by the classical evolution $a=a(t)$.

We now display the relationship between this linear development giving rise
to time through $a=a(t)$
%and the concept of the classical trajectory 
with the
linear development in the momentum transfer to the massive detector which 
leads to the concept of its proper time $\tau$ and its classical trajectory $z=z(t)$.

This second approximation is defined by developing 
to first order in $\Delta m$ and 
%to first order in
the momentum transfer $k$ 
the difference of the energies $\Om(t)$ 
in eq. (\ref{diffta1}) exactly as we developed the difference of momenta $
\pi(a, n_f) - \pi(a, n_i)$ in the energy change. One gets
\ba
\int_t^{t_2} \left[ \Om( M, p -k , a(t') ) - \Om(m, p, a(t') ) \right] dt' &=&  
 \int_t^{t_2} { dt'  \ m \over \Om(m, p, a(t')) } \left[ \Delta m - { k \over a(t')^2} 
 {  p \over m} \right] 
\nonumber\\ &=&
\int_\tau^{\tau_2} d\tau' \left[ \Delta m - { k \over a(\tau')^2}  {  p 
\over m} \right] 
%\int_{\eta}^{\eta_2} d\eta 
%\left( 1 + { p k \over a(\eta') \Om(m, p , a(\eta')) |k| } \right)
\label{difta11}
\ea
where we have introduced the proper time (of the particle of momentum $p$)
defined by $\tau = \partial_m S$ where $S$ 
is the Hamilton-Jacobi action equal to $\int dt \ \Om(m, p, a)$.

Both terms are easily interpreted. 
%\ee
%Indeed the proper time $\tau$ is defined by $\tau = \partial_m S$ where $S$ 
%is the Hamilton-Jacobi action.
The first term corresponds to the phase of the excitation energy $ \Delta m $
evaluated 
along the classical trajectory, hence parametrized by $\Delta \tau$.
Similarly, the second term is
\be
%|k| \int_{a}^{a_2}  { da'  \over G \pi(a', n_i) } 
%( 1 + { p k \over a' |k| \Om(m, p , a') } ) 
%|k| \int_{\eta}^{\eta_2} d\eta \left( 1 +  { v(\eta , p ,m )
%\over a(\eta')}{ k \over |k|}\right)
k  \int_t^{t_2} { dt' \ p 
\over a(t')^2 \Om(m, p, a(t')) } = k  \partial_p \int_t^{t_2} 
 dt' \ \Om(m, p, a(t'))
%=  \int_t^{t_2} { dt'  v (t , p ,m ) } = 
= k \Delta z(t_2, t )
%\over \Om(m, p, a(t') }
\label{etav}
\ee
since by definition the distance is given by $\Delta z =  \partial_k S \vert_{k=0}$.
Then, by grouping together the term in $|k|$ in eq. (\ref{diffta1})
and this latter term one gets
%where the we have used the velocity vector 
%of a massive particle of momentum $p$ defined in eq. (\ref{velo}).
%
%By defining the vector displacement as $\partial_{\eta} 
%\Delta z = v(\eta )/ a(\eta )$,
%we see that eq. (\ref{etav}) can be written as
\be
|k| \Delta \eta( t_2, t ) - k \Delta z ( t_2, t ; m , p)
\label{wfphot}
\ee
This corresponds, as it should be, to the $k$ component of
the phase of the Green function  
of a massless quantum created at $a$ 
%and destroyed at $a_2$
by the heavy
detector which follows the classical trajectory 
$\partial_{\eta} \Delta z = a(\eta ) v(\eta )$, where $v$ is 
defined in eq. (\ref{velo}).
Furthermore, to this order
in $\Delta m$ and $k$, the {\it norm} of the amplitude is
 equal to $dt/ \Om( m, p , a(t)) = d\tau$,
thanks to the $a$ dependence of $\bra{n_f} \tilde H_{int}(a) \ket{n_i}$,
see eq (\ref{adep}).

Collecting the results, we have for the amputated amplitude 
\ba
{\cal{B}}_{amp}^{n_i}(a_2, a_1 ; p, k) &=& 
- i  
\int_{a_1}^{a_2} {a da \over G  \pi(a, n_i)} \ 
\bra{n_f} \tilde H_{int}(a) \ket{n_i} 
\ e^{i \int_a^{a_2} [\pi(a', n_f) - \pi(a', n_i) ] da'}
\nonumber\\
&=& - ig \int_{t_1}^{t_2} {  mdt  \ \sqrt{n_m ( n_M+1 ) n_k }
\over   \Om(m, p, a(t)) } \  e^{i  \int_{t}^{t_2} 
\left[ \Om( M, p -k , t' ) - \Om(m, p,t' ) \right] dt' } { e^{i |k| (\eta_2 - \eta)} 
\over \sqrt{ |k|}}
\nonumber\\
&=& - ig \sqrt{n_m ( n_M+1 ) n_k } \ \int_{\tau_1}^{\tau_2} {  d \tau } \
e^{i  \Delta m (\tau_2 - \tau ) } 
{e^{i |k| (\eta_2 - \eta) - k \Delta z} \over   \sqrt{|k|}}
\label{sucess}
\ea
%We emphasize this point, to first order in $k$, the mass $m$ and the momentum
%$p$ of the heavy detector disappears in the favor of the properties of 
%the sole classical trajectory through the use of the dispersion relation, see 
%eq. (\ref{velo}).
The parallelism between the development in the energy change in 
eq. (\ref{diffta1})
and the one in the momentum transfer in eq. (\ref{difta11})
% we did  in eq. (\ref{diffta1}) (and in eqs. (\ref{difftata}, \ref{better2}))
 is manifest.
% what happens in eqs. (\ref{difftata}, \ref{better2}):
In both cases,
by making a first order 
%development in the energy (or momentum) 
change in the
WKB kernel (i.e. the product of two wave functions)
 of the heavy system, i.e. the universe and the atom respectively,
one recovers the transition amplitude evaluated at the BFA wherein the
heavy system is described by a single classical trajectory.

Indeed, in the case of gravity, the first order change allows the replacement 
of the quantum characteristics of the wave function of the 
universe, such as $\Lambda$ and $ n_i$,
in favor of the properties of
the sole classical trajectory of the universe $a(t)$.
This replacement is effected by means of the
Hamilton-Jacobi equation $dt/da  =  \partial_{E_M} \pi (a, n_i) =
 a / G \pi (a, n_i) $,  i.e. the dispersion relation for the universe.

Similarly, in
 the heavy particle case,  the first order change in the momentum transfer
allows the replacement of the momentum 
and the energy of the particle by its classical trajectory. 
This is effected by means of the
Hamilton-Jacobi equations of the particle,
$d\tau / dt = \partial_M \Om( M, p , t)$ and $dz / dt = \partial_p \Om( M, p , t)$
%and  $d\tau / dz$
%In both cases, if one stops the expansion to first order, 
%one recovers the amplitude evaluated at the BFA.

This parallelism proves that the universe behaves like an heavy system, at 
in the WKB approximation. We shall briefly discuss in the next Section 
what happens when one leaves the WKB approximation and considers
``negative'' energy solutions.

The other lesson is that
 we have recovered the usual spatial dependence of the 
transition amplitudes from the $k$ dependence of the
action of gravity $\int da \ \pi_a$  through the dispersion relation, eq. (\ref{disp}).
Therefore, gravity acts also like a reservoir in delivering spatial 
displacements. See \cite{wdwgf} for a systematic comparison
between the behaviour of an heat reservoir delivering a temperature
and the one of gravity delivering time lapses.
%Then upon considering wave packets in $k$, the three dimensional local
%character of field theory can be recovered from the correlations among
% homogeneous solutions of the WDW equation. 

We conclude by noting that there is, by 
construction, a {\it single} inertial frame (introduced in this formalism
by first order variations with respect to momentum transfers).
This frame is singled out by the fact that we are working with
solutions of the WDW equation. Indeed, the homogeneous part 
of the three-momentum constraint implies that the total 
momentum of matter vanishes. This is turn defines 
what are the conserved momenta. Therefore both spatial 
and temporal inertia are defined {\it through gravity} by the 
macroscopic matter content of the universe. 
In fact, it is the macroscopic character of the sources
of gravity which leads to the concept of background 
defined by a first order expansion in $1/n$ in eq. (\ref{diffta1}).
Isn't it Mach's principle ?

\section{The non-linearities induced by quantum gravity}

In this Section, we evaluate the corrections to the various approximations
 used so far.

We have used four such approximations. Therefore there will be four types
of corrections and our task is to  isolate the dominant ones.

First we have considered that the spread in energy in the initial matter state
vanishes. Secondly, we have developed the expressions to first order
in the energy change characterizing the transition, see eq. (\ref{difftata}).
Thirdly we have the corrections to the WKB approximation used to
evaluate the free kernels, see eq. (\ref{M16'}). Finally we have considered
amplitudes to first order in $g$ only, see eq. (\ref{pii}).

We start by discussing the first type of non-linearities, the ones
associated with the energy spread of the initial state.
To this end, it is appropriate to deal with the probability of transition,
see eq. (\ref{GR1}).
Since the amputated ${\cal{B}}_{amp}^{n_i}(a_2, a_1)$, eq. (\ref{better2}),
defines the amplitude of {\it probability} in quantum gravity,
the probability of emitting a massless quantum of momentum
 $k$ is given by 
\be
{\cal{P}}^{n_i}(k; a_2, a_1) = 
\vert {\cal{B}}_{amp}^{n_i}(a_2, a_1 ; k) \vert^2
\label{prob'}
\ee
We want first to determine to what extend the Golden 
Rule emerges from this probability.
To this end, we consider the case of an
 almost de Sitter space, i.e. the solution of eq. (\ref{modwdw}) 
in which  the cosmological term
in $\Lambda a^4$ dominates the matter terms.
Then the rate of transition $R$ 
is determined by the Hubble constant $h$ only.

To obtain the total probability of transition,
as in eq. (\ref{GR1}), we integrate over all $k$ 
\ba
{\cal{P}}^{n_i}(a_2, a_1) &=& \int_0^\infty { dk \over 2 \pi} 
\vert {\cal{B}}_{amp}^{n_i}(a_2, a_1 ; k) \vert^2  
\nonumber\\
 &=&  
%R  \int_{a_1}^{a_2} {a da \over G \pi(a, n_i) } = 
 R \  n_m (n_M + 1 ) \ \Delta t (a_2, a_1; n_M, n_m, n_\gamma = 0 )
\label{CALP}
\ea
where we have used eq. (\ref{better2}) to evaluate the rate $R$.
We see that the probability grows linearly
with respect to the proper time evaluated in the 
(almost de Sitter) background characterized by 
$n_M, n_m$ and $ n_\gamma = 0$ quanta.

Therefore it is instructive to consider the probability of transition when the 
initial matter state is described by a superposition of states 
characterized by $n_M, n_m$ with amplitudes $c_{n_M, n_m}$, 
see eq. (\ref{sump}).
In the present case the mean probability is given by
\ba
\bk{{\cal{P}}(a_2, a_1)} &=& \Sigma_{n_i} | c_{n_i} |^2 {\cal{P}}^{n_i}(a_2, a_1) 
\nonumber\\
 &=&  R \
 \Sigma_{n_i} | c_{n_i} |^2 ( n_M+1) n_m 
%\int_{a_1}^{a_2} {a da \over G \pi(a, n_i) }
%= g^2 R  \ \Sigma_{n_i} | c_{n_i} |^2 
\Delta t (a_2, a_1; n_M, n_m, n_\gamma =0 )
\label{CCALP}
\ea
The time lapse no longer factorizes since it
depends parametrically of the matter content. 
The origin of this {\it non-linearity} is that the occupation numbers $n_M, n_m$ play 
a double role. Not only do they give the norm of the probability as at the BFA,
 see the overall factor $ ( n_M+1) n_m $, but they also 
determine parametrically the lapse of time 
$\Delta t (a_2, a_1; n_M, n_m, n_\gamma =0 )$
since the latter is {\it defined} upon evaluating 
the phase of the amplitude itself, see eq. 
(\ref{difftata}).

%We do not claim that this is a big effect in the present de Sitter space.
%As a matter of principle it is however important to determine
%how the quantum dynamical character of gravity modifies
% matter transition amplitudes evaluated at the BFA.
%.grows linearly
%with respect to the mean proper time. Therefore w

When the 
spread in $n_i$ is negligible as compared to the 
mean occupation numbers $\bar n_i$, 
the {\it mean} probability can be correctly approximated
by the probability evaluated 
in the {\it mean} universe, see \cite{wdwgf} for
an evaluation of the corrections.
 In this case only, one fully recovers the concept
of transition amplitudes evaluated in a {\it single} gravitational background
which is completely insensitive to the individual quantum matter processes.

On the contrary, 
when the spread is big or comparable to the mean number, the
validity of describing gravity as a unique background looses sense.
However, since each component engenders its own but well defined time
 parametrization,
the impossibility of defining 
 a single background is free of physical consequences.
Indeed, remote configurations never interfere.
The reason for this is very simple.
% lies on the following remark. 
In order to connect
quantum mechanically two matter states characterized by $n_1$ and $n_2$ quanta,
one needs a interaction hamiltonian (or a time ordered product of this hamiltonian)
which  possesses non vanishing matrix elements between these states.
Therefore, given a certain matter hamiltonian, two remote initial states
and a certain lapse of time,
one easily checks whether or not the two wave functions might interfere.

In our case, since $N = n_M + n_m$ is strictly conserved, initial states made 
of superposition of various $N$ cannot interfere. Therefore their evolution 
is completely independent of each other. Thus it is of no use 
to consider a superposition of states with different $N$. In particular,
it is of no use to consider a superposition which leads to a 
``quasi-classical state''\cite{ortiz} describing a single background. 
%In this we disagree with \cite{ortiz}.
%this is valid for the time parametrization. 
The situation is different for the radiation field since the number of photons 
which can be produced by intercations with massive detectors does 
fluctuate dynamically. In that case, it is mandatory to consider
 interferences among neighbouring components in the wave function.

The second type of corrections are associate with the non-linear terms
in the energy change which are present in the norm and the phase of 
the amplitude ${\cal{B}}$, see eq. (\ref{difftata}).

The algebra to obtain them is very similar to the one of Section 5 of \cite{wdwgf}.
Therefore we shall just comment the results. The
non linear corrections to eq. (\ref{difftata}) are given by
\ba
\int_{a}^{a_2} \left[ \pi(a', n_M + 1, n_m - 1, n_\gamma + 1) -
\pi(a', n_M, n_m, n_\gamma)
\right] da' =\quad \quad\quad \quad\quad \quad \quad &&
\nonumber\\
\int_{a}^{a_2} \left[ { ( a' \Delta m  + k )\over G \pi(a', n_i) } 
 \left( 1 - { ( a' \Delta m  + k ) \over G \pi^2(a', n_i) } \right)\right] da'
\simeq \quad \quad \quad \quad \quad \quad \quad &&
\nonumber\\
 -  \left[  \Delta m \Delta t( a_2, a ; n_i) + k \Delta \eta( a_2, a ; n_i) \right] 
\left[ 1 - {( a \Delta m  + k )\over G \pi^2(a, n_i) }
\left( 1 +  \Delta t
%( a_2, a ; n_M, n_\gamma)
({ \dot a \over a } -
 {2 \dot \pi
%(a, n_M, n_\gamma)
\over \pi
%(a, n_M, n_\gamma)
}) \right) \right] \quad \quad  &&
\label{C1'}
\ea
In the third line we have replaced the integral of the
correction by the value of the integrand at $a$ times the interval
plus the linear dependence in $ \Delta t$.
%This is valid if $a / \pi^2$ is slowly varying.
To estimate these terms, one should specify the dominant
matter content of the universe.
When the universe is matter dominated, the first correction term is
\be
{( a \Delta m + k )\over G \pi^2(a, n_i)}
\simeq { (\Delta m + k/ a )\over 2 ( M n_M + m n_m + 2 k n_\gamma /a) }  
\simeq { \Delta m \over 2 m (n_M + n_m )}
\label{C2}
\ee
It is independent of $G$, $\hbar$ and $a$. It depends only on the
ration of the matter change by the total energy in the universe.
This is what guarantees the validity of the background field approximation:
that the sources of gravity be heavy when compared to 
the transition upon examination. 
%This is our third result.

As in Section 5 of \cite{wdwgf}, the origin of this term arises
from our choice to develop
eq. (\ref{C1'}) around the initial occupation numbers $n_i$.
Indeed,
had one developed around the ``mean'' values $n_M + 1/2, n_m - 1/2 , 
n_\gamma + 1/2$,
which characterize the transition under examination,
 this dominant term would not be present.
Its meaning is therefore that the ``background'', i.e. the best 
gravitational fit around which one develops eq. (\ref{C1'}),
 is determined by the 
transition under examination. Furthermore, the initial and 
the final number of quanta play a symmetrical role in defining 
this background.
Therefore we can say that the geometry is ``post-selected''\cite{ah}
by the transition under examination.
Whether or not this quantum specification of the background
leads to modifications of dynamical processes big enough to
invalidate the treatment based on single insensitive background
remains to be seen. The reader 
may consult refs. \cite{THooft2}\cite{emp}\cite{fe}\cite{itz}\cite{recV}
where this point of view is advocated.

Very important also is the fact that the correction terms scale with the
energy change $\Delta m$ and not with the two masses $M$ and $m$.
Therefore, the corrections to the probability cannot
be expressed in terms of corrections of each propagator since
the latters are controlled by the masses. Thus the probability 
of transition cannot be written as the convolution of three
 Wightman functions as it is done usually, see eq. (\ref{GR1}).
This is another manifestation of the non-linearities of quantum gravity.

Finally, we stress that the amplitude of transition ${\cal{B}}$ is given
in term of an integral over the radius $a$ at which the interaction acts.
Therefore, the non-linear corrections to the BFA, i.e. the non linear
terms in eq. (\ref{C1'}),
% which determine the 
%quantum specification of the background, 
are summed at the
level of the amplitude.
This means that the ``post-selected'' background is summed as well.
It is not only sensitive to the initial and 
the final number of quanta, it depends also on the dummy radius $a$ 
at which the
interaction occurs.
This fact reinforces the impossibility of dealing with a {\it single}
background once the dynamical character of gravity is taken at the 
quantum level. 
% into account 

The third type of corrections are associated with the WKB
approximation used for the free kernels $K$, see eq. (\ref{M16'}).
These corrections have been discussed in Section 5 of \cite{wdwgf}.
We just recall that the are irrelevant 
when the Hubble radius satisfies $\dot a / a << m$. Indeed, in that case,
 they are smaller than the 
former corrections by a factor of 
$1/(n_M + n_m)$ for matter dominated
universes.
% when the Hubble radius satisfies $\dot a / a << m$.

The fourth type of corrections is much more interesting.
It has to do with amplitudes evaluated at higher orders in $g$.
Up to now we have only considered amplitudes which are linear 
in the hamiltonian $H_{int}$, see eq. (\ref{i1}).

To order $g^2$, one finds now two contributions.
The first contribution arises from the linear term of eq. (\ref{pii}) acting twice
whereas the second one comes from the quadratic term of this 
equation acting once.
When using the WKB approximation to describe the propagation of
gravity, the first contribution leads back to the {\it time}-ordered
product of the interaction hamiltonian. 
Indeed, the monotonic character
of the expansion allows the replacement of the $a$-ordered product
to a time-ordered one.
The second contribution to order $g^2$ can be conceived as a 
``contact''  interaction induced by gravity, i.e.
the large mass limit of a $Z$ graph. For
macroscopic
universes, it is negligible when compared to the first contribution
since it scales like $
\Delta m / (M n_M + m n_m )$. It has therefore
 the same magnitude of the corrections due to the non-linearity
of the phase, see eq. (\ref{C1'}). 

To conclude this analysis of the non-linearities
induced by the dynamical character of gravity,
we develop the remark made after eq. (\ref{sucess}).
This remark mentioned the strict parallelism between the
behaviour of gravity and that of an heavy
relativistic atom.
This analogy was established at the WKB approximation only.
What happens when one abandons this approximation ?
This raises the question of the role of negative energy states
and the possibility of creating pairs of heavy systems.

The modifications induced by this abandonment have been 
discussed and evaluated in \cite{suh} in the relativistic
particle case. In that case, when dealing with
non-vanishing amplitudes to create pairs of particles,
 ``wave functions'', i.e. the solutions
of the Klein-Gordon equation, no longer possess
a probabilistic interpretation.
However, in spite of these non-vanishing pair
creation amplitudes, it was shown that the use of the
{\it in-out} propagator for describing the propagation of 
the relativistic atom allows the study 
of the emission amplitudes (bremstrahlung) in usual terms,
at least at the tree level.
This indicates that there is no dramatic consequence in 
having lost the statistical interpretation of the
solutions of the Klein-Gordon equation.
We are inclined to believe that the situation might be similar 
for the solutions of the Wheeler De Witt 
equation.

\vskip .5 truecm
{\bf Acknowlegdments    }

\noindent
I thank Robert Brout for very useful comments, remarks and suggestions.

\vskip .3 truecm
 \section{Appendix A.
\newline
The transition amplitudes in de Sitter space}

In this Appendix we compute the transition amplitudes 
in the de Sitter background. 
In particular we emphasize the relationship between the 
exponentially growing Doppler effect  
%event horizon 
and the cosmological temperature\cite{HH}.

%In de Sitter space,
%In that background,
%When the universe is de Sitter dominated,
In a flat expanding de Sitter background, the scale factor $a$
satisfies $\dot a / a = h = \Lambda^{1/2} = constant$.
Therefore the lapse of
conformal time lapse is 
\be
\Delta \eta (t_2, t)= - e^{-h t_2} + e^{-h t}
\label{etads}
\ee
This exponential Doppler shift relating conformal and proper time
signals the presence of a static event horizon whose surface gravity
is equal to the inverse Hubble radius $h$\cite{wald}. 

In that background, the amplitude $B$ of spontaneous
excitation is, see eq. (\ref{calB}),
\be
{B}^{n_i}(t_2, t_1 ;k) = -ig \ \bra{n_f}  \psi_M^\dagger \psi_m  \phi \ket{n_i}
\
\int_{t_1}^{t_2} dt \  e^{-i \Delta m (t_2 - t) }
e^{i k (e^{-h t_2} - e^{-h t})} 
%\times \left( e^{- im(t_2 - t_1)} \right)
\label{calB2'}
\ee
We have dropped the irrelevant phase of the free propagation of the initial state.
This amplitude is identical to the one of an accelerated system in
 Minkowski vacuum
whose excitation energy is $\Delta m$ and whose acceleration is $h$.
The proper time of the inertial atom $t$ plays the role
of the proper time of the accelerated atom and the conformal time plays the 
role of the Minkowski time, see \cite{unruh}\cite{BD}\cite{GO}.
%variable

In the limit $t_1 \to - \infty$, $t_2 \to \infty$, this
amplitude can be computed analytically. It is given
by 
\ba
%\lim_{t_1 \to - \infty, t_2 \to \infty} {B}^{n_i}(t_2, t_1 ;k) &=&
{B}^{n_i}(k) 
%\nonumber\\
&=& -ig \ \bra{n_f}  \psi_M^\dagger \psi_m  \phi \ket{n_i}
\ \Gamma({-i \Delta m \over h}) \left({ k h \over \Delta m } \right)^{- i \Delta m / h}
{ e^{-\pi  \Delta m / 2h}} \quad 
\label{calb3}
\ea
where $\Gamma$ is the Euler gamma function\cite{BD}\cite{GO}.

But what we really nead is the ratio of $B$ to the inverse
transition where one massive state $M$ disintegrates into
a lighter one plus a massless $\phi$ quantum.
This inverse transition is given by complex conjugating and 
replacing $k$ by $-k$ in
eq. (\ref{calB2'}): 
\be
A^{\tilde n_i}(t_2, t_1 ;k)
= \left( B^{n_i}(t_2, t_1 ; -k) \right)^*
%\times \left( e^{-i \Delta m (t_2 - t_1)} \right)
\label{calA}
\ee
We have introduced the
symbol $\tilde n_i = ( n_M +1 , n_m - 1, n_\gamma)$ to designate
the initial numbers of quanta appearing in $A$, taking into account
the Bose statistic. 
%The overall phase $ e^{-i \Delta m (t_2 - t_1)}$ 
%plays no role in determining rate but ...
By using the determination of the logarithm of $k$ in eq. (\ref{calb3}),
one finds that the
ratio of the square of the norms satisfies the
Boltzmanian law
\be
\lim_{t_1 \to - \infty, t_2 \to \infty}
{ |B^{n_i}(t_2, t_1 ; k)|^2 \over |A^{\tilde n_i}(t_2, t_1 ;k)|^2}
= {|B^{n_i}( k)|^2 \over |A^{\tilde n_i}(k)|^2}
= e^{- 2 \pi \Delta m / h }
\label{rat}
\ee
This ratio indicates that, in the vacuum of the $\phi$ field,
 the mean numbers of massive
fields satisfy, at equilibrium, 
\be
{\bar  n_M \over \bar n_m} =
{ |B^{n_i}(t_2, t_1 ; k) |^2 \over |A^{\tilde n_i}(t_2, t_1 ;k)|^2}
= e^{- 2 \pi \Delta m /h }
\label{rat2}
\ee
Thus, in de Sitter space, inertial massive systems
experience a cosmological temperature given by $h / 2 \pi$\cite{HH}.

We recall that this exponential ratio can be understood in terms
of the saddle point analysis\cite{PB2}\cite{GO}. We present this analysis
to introduce the concept of resonance which designates where the 
transition occurs.
% and which leads to non vanishing rates. 
Furthermore, the classically forbidden transition described by the
amplitude $B$ leads
to a saddle point which possesses an imaginary part associated with the
presence of a static horizon. Both aspects are also found in 
quantum cosmology, see Appendix B.

The saddle points
for $A$ and $B$ are given by the solution
\be
\Delta m = \pm { k  h} \ e^{- h t_{\pm}}
\label{sps}
\ee
where the $+$ sign is for the amplitude $A$.
One finds
\be
t_- = t_ + + i {\pi /h }
\label{sps2}
\ee
where $h t_+ = \ln( h k / \Delta m ) $ is real. It determines at which proper time $t$ 
the conformal 
Doppler shifted frequency $k/a(t)$ equals (i.e. resonates with)
the proper frequency $\Delta m$.

Instead, the imaginary part of $t_-$, $i {\pi /h }$, is independent of $k$.
This is crucial and related to the presence of a static
event horizon. This event horizon requires that
 the periodicity in Im$(t)$ be equal to $2 \pi /h$ in both
the euclidian continuation of the metric and in the Green functions of the
 radiation field in order to have respectively, no conical singularity and regular
energy density at the horizon. 

Furthermore, the fact that the imaginary part of $t_-$
is independent of $k$ leads immediately that
for $h k$ belonging to the interval $[ \Delta m e^{ h t_1}, \Delta m e^{ h t_2}] $, one 
obtains, at the saddle point approximation, 
\be
 {B^{n_i}(t_2, t_1 ; k)
\over A^{\tilde n_i}(t_2, t_1 ;k) } 
= e^{- \pi \Delta m / h } \times \mbox{ phase}
\label{baba}
\ee
When $h k$ does not belong to that interval, the amplitudes are dominate by transient
effects and do not lead to constant 
rates of transition, see \cite{GO}.

The saddle point approximation offers therefore a 
description of the transition which reveals the mechanisms
into action. This is why we shall also apply it 
directly to the amplitudes evaluated in quantum cosmology.
%This method has been applied
The results we shall find are in strict analogy
with the results of \cite{rec}
wherein the saddle point approximation
was used to analyse the effects of the recoils
of an accelerated two level atom induced by the 
Unruh effect.

\section{Appendix B.
%\newline
The saddle point evaluation of
the transition amplitudes in quantum cosmology }

In this Appendix we shall apply the same saddle point 
techniques to the amplitudes in quantum cosmology.
The interest of this treatment is that it does not require to have
defined
 a background nor does it lead to the notion of a background. 
This is in contradistinction to the BFA wherein a background 
is {\it a priori} given,  as well as to what we did in 
Sections 3 and 4 where we defined a background by
 {\it first} developing the expressions in powers of 
the energy changes.
% in order to compare them with the corresponding
% amplitudes evaluated in the BFA. 

Here, as in \cite{rec}\cite{recmir}, we consider the stationary phase condition
applied to the amplitude expressed in terms of the dynamical variable itself. 
We recall that to first order in matter interaction and to the WKB
approximation for the propagation of gravity, the 
amplitude ${\cal{B}}_{amp}^{n_i}(a_2, a_1 ; k)$ 
%and ${\cal{A}}^{\tilde n_i}(a_2, a_1 ; k)$ are 
is given by, see eq. (\ref{better2}),
\ba
\int_{a_1}^{a_2} {a da \over G \sqrt{ \pi(a, n_f) \pi(a, n_i)}}
\ e^{ i \int_a^{a_2} [\pi(a', n_f) - \pi(a', n_i) ] da'}
\label{baba'}
\ea
%where the $+$ sign is for the amplitude ${\cal{B}}^{n_i}$, see eqs. (\ref{}\ref{}\ref{}).
Thus the stationary condition gives simply
\be
\pi(a^*, n_f) = \pi(a^*, n_i)
\label{pipi}
\ee
where the momenta are evaluate without the interaction hamiltonian, see eq. (\ref{pii'}).
Thus the saddle radius $a^*$ is such that the two universes,
characterized respectively by $n_i$ and $n_f$ quanta, resonate. This is 
 exactly like in eq. (\ref{sps}).
Furthermore, this condition gives, to all order in $G$,
\be
E_{n_f} ( a^*) = E_{n_i} (a^*)
\label{EE}
\ee
That is the two universes resonate when their matter energy coincide.
This is as it should be since the momentum of gravity is entirely
specified by $a$ and $E_{n_i}$, see eq. (\ref{pii'}).
% in \cite{wdwgf}.

Thus for the amplitudes ${\cal{B}}^{n_i}(a_2, a_1 ; k)$ 
and ${\cal{A}}^{\tilde n_i}(a_2, a_1 ; k)$, we get, instead of eq. (\ref{sps}),
\be
\Delta m = \mp {k \over a^*}
\label{overa}
\ee
Therefore, for the classically permitted transition ${\cal{A}}$, the saddle 
radius $a^*$
is such that the Doppler shifted  conformal energy equals the energy gap 
$\Delta m$.
On the contrary, for the forbidden transition ${\cal{B}}$, the saddle $a^*$ 
is negative and not classically accessible.
Both are related to the solutions of eq. (\ref{sps})
if one {\it introduces} the classical trajectory of the
background $a= a(t) = e^{-ht}/h$.

How the non classical 
 saddle point condition is related to
euclidian general relativity, the  entropy of gravity 
 as derived in \cite{gh},  and the change in the area induced by the 
transition under investigation are interesting questions presently under 
examination.

\end{document}